\documentclass[11pt]{article}
\usepackage{amsmath}
\usepackage{graphicx}
\usepackage{enumerate}
\usepackage{natbib}
\usepackage{url} % not crucial - just used below for the URL

\usepackage{xr} % Necessary to reference to Fig and Tab in
\newcommand{\blind}{1}

\usepackage{amsthm}
\theoremstyle{definition} % Non-italic type theorem
\newtheorem{algo}{Algorithm} % For Algorithm sections...

\usepackage{mathtools} % $\overline{text}
\usepackage[usenames,dvipsnames]{color}

\usepackage{enumerate} % Package useful to set symbol in itemize-like
\usepackage[framemethod=TikZ]{mdframed}

\usepackage[labelsep=endash]{caption}
\newcounter{Algocount}
\newcounter{tempFigure}

\usepackage{array}
\newcolumntype{L}{>{\arraybackslash}m{.92\textwidth}}

\usepackage{bbm} % To get the indicator function via \mathbbm{1}

\newcommand{\E}{\mathbb{E}}
\newcommand{\V}{\mathbb{V}}

\usepackage{amsfonts} % \mathbb{set} where set is R, N, Z, etc for  |R, |N, etc.
\usepackage{eurosym}
\usepackage{natbib}
\usepackage{epsf}
\usepackage{graphicx}
\usepackage{rotating}
\usepackage{tabularx}

\usepackage{arydshln}

\usepackage[latin1]{inputenc}

\newcommand{\beq}{\begin{equation}}
\newcommand{\eeq}{\end{equation}}

\numberwithin{equation}{section} % Equation numbering preceded by Section number
\newcommand{\rit}{{\rm I\!R}}

\newcommand{\ttf}[1]{{\ttfamily{#1}}}

\newcommand{\tr}[1]{\mathrm{tr}\left({#1}\right)}

\newcommand{\kron}{\otimes}
\newcommand{\diag}{\mathrm{diag}}

\newcommand{\N}[2]{{\cal N}\left(#1,#2\right)}
\newcommand{\Nc}{{\cal N}}
\newcommand{\G}[2]{{\cal G }\left(#1,#2\right)}
\newcommand{\D}{{{\cal D}}}

\newcommand{\Exp}{{\mathrm{Exp}}}

\renewcommand{\Vec}[1]{\mathrm{vec}{(#1)}}

\DeclareMathAlphabet\mathbfcal{OMS}{cmsy}{b}{n}

\begin{document}

\def\spacingset#1{\renewcommand{\baselinestretch}%
{#1}\small\normalsize} \spacingset{1}

%%%%%%%%%%%%%%%%%%%%%%%%%%%%%%%%%%%%%%%%%%%%%%%%%%%%%%%%%%%%%%%%%%%%%%%%%%%%%%

\title{\bf Fast Bayesian Inference in Nonparametric
  Double Additive Location-Scale Models With Right- and
  Interval-Censored Data
  % \title{\bf Inference Based on Laplace Approximations in Nonparametric
  % Double Additive Location-Scale Models With Right- and
  % Interval-Censored Data
}
\date{January 28, 2020}
\if1\blind
{
  \author{Philippe Lambert\footnote{
    Institut de Recherche en Sciences Sociales (IRSS),
    Universit\'e de Li\`ege, Belgium. Email: p.lambert@uliege.be}
    \footnote{Institut de Statistique, Biostatistique et Sciences Actuarielles
    (ISBA), Université catholique de Louvain, Belgium.}}
  \maketitle
} \fi

\if0\blind
{
  \bigskip
  \bigskip
  \bigskip
  \begin{center}
    {\LARGE\bf {\bf Fast Bayesian Inference in Nonparametric
        Double Additive Location-Scale Models With Right- and
        Interval-Censored Data}
\end{center}
  \medskip
} \fi

%\bigskip
\begin{abstract}
  Penalized B-splines are routinely used in additive models to
  describe smooth changes in a response with quantitative
  covariates. It is typically done through the conditional mean in the
  exponential family using generalized additive models with an
  indirect impact on other conditional moments.  Another common
  strategy consists in focussing on several low-order conditional
  moments, leaving the complete conditional distribution unspecified.
  Alternatively, a multi-parameter distribution could be assumed for
  the response with several of its parameters jointly regressed on
  covariates using additive expressions.

  Our work can be connected to the latter proposal for a right- or
  interval-censored continuous response with a highly flexible and
  smooth nonparametric density.  We focus on location-scale models
  with additive terms in the conditional mean and standard deviation.
  Starting from recent results in the Bayesian framework, we propose a
  quickly converging algorithm to select penalty parameters from their
  marginal posteriors. It relies on Laplace approximations to the
  conditional posterior of the spline parameters.  Simulations suggest
  that the so-obtained estimators own excellent frequentist properties
  and increase efficiency as compared to approaches with a working
  Gaussian hypothesis. We illustrate the methodology with the analysis
  of imprecisely measured income data.
\end{abstract}

\noindent%

\section{Introduction}
Additive models are flexible alternatives to the classical linear
regression model to describe in a flexible way the effect of
quantitative covariates on various aspects of a response
distribution. Early proposals focussed on the conditional mean with
limited assumptions on the conditional distribution of the response
\citep{BreimanFriedman:1985}. That idea was used to extend generalized
linear models \citep[GLM,][]{Neld:Wedd:1972} and the analysis of
nonnormal data (such as counts or proportions) in the framework of the
exponential family of distributions: additive terms enter the GLM
linear predictor (connecting covariates to a pre-specified function of
the conditional mean) for a fixed value of the dispersion parameter,
yielding generalized additive models (GAM)
\citep{Hast:Tibs:1986,Hast:Tibs:1990,wood2017generalized}.
Further extensions are possible by enabling covariates to also affect
other aspects of the response distribution such as dispersion,
skewness and kurtosis, see \citet{Lambert1999} for early work on this
with the four parameters of the stable distribution simultaneously
modelled
and \citet{Rigby2005a} for an extension to a large choice of parametric
distributions.  \citet[Chap.\,11]{Lee:Nelder:Pawitan:2006} and
\citet{Gijbels:Prosdocimi2012} considered joint additive models for
location and dispersion within, respectively, the exponential and the
double-exponential families of
distributions, while \citet{Croux:Gijbels:Prosdocimi:2012} relied on a
(robustified) extended quasi-likelihood method.

Our paper will focus on double additive models for the conditional
mean and standard deviation in location-scale models with a
nonparametric error distribution. The response will be assumed
continuous and possibly subject to right or interval censoring.
Nonparametric inference from censored data in location-scale models
has been investigated by many authors, see
e.g.\ \citet{Fan:Gijbels:1994} for early work using local polynomials
and \citet{Heuchenne:IVK:2010} with the references therein for some
more recent work. These methods typically focus on the estimation of
the conditional location and can only handle the estimation of the
smooth effects of a very limited number of covariates. Additive models
based on P-splines \citep{Eilers1996,Lang2004} are preferred here for
their excellent properties \citep{Eilers2010} and the possibility to
handle a large number of additive terms.  They are used to specify the
joint effect of covariates on location and dispersion in the framework
of the location-scale model, see Section
\ref{AdditiveLocationScale:Sec}. A nonparametric error distribution
with an underlying smooth hazard function and fixed moments will be
assumed for the standardized error term, see Section
\ref{NPDensityEstimation:Sec}.
In the absence of right censoring, a location-scale model with a small
number of additive terms and a quartile-constrained error density
(instead of the hazard here) was considered in \citet{Lambert2013} to
analyse interval-censored data, with inference relying on a
numerically demanding MCMC algorithm.
We show how Laplace
approximations to the conditional posterior (of blocks) of spline
parameters can be combined to bring fast and reliable estimation of
the additive terms in the location and dispersion models, and provide
a smooth estimate of the underlying error hazard function under moment
constraints. These approximations are the cornerstones in the
derivation of the marginal posteriors for the penalty parameters and
smoothness selection, see Sections \ref{MarginalPostLambda:Sec} and
\ref{SelectionPenaltyParmTau:Sec}. The resulting
estimation procedures are motivated using Bayesian arguments and shown
to own excellent frequentist properties, see Section
\ref{SimulationStudy:Sec} and Supplementary Material \ref{Appendix:C}. They are
extremely fast and can handle a large number of additive terms within
a few seconds even with pure R code. The methodology is illustrated in
Section \ref{Applications:Sec} with
the analysis of right- and interval-censored income data in a
survey. We conclude the paper with a discussion in Section
\ref{Discussion:Sec}.

%%%
\section{Additive location-scale  model} \label{AdditiveLocationScale:Sec}

Consider a vector $(Y,\mathbf{z},\mathbf{x})$ where $Y$ is a univariate
continuous response, $\mathbf{z}$ a $p-$vector of categorical covariates,
and $\mathbf{x}$ a $J-$vector of quantitative covariates.
The response could be subject to right censoring, in which case one
only observes $(T,\Delta)$, where $T=\min\{Y,C\}$, $\Delta=I(Y\leq C)$
and $C$ denotes the right censoring value that we shall assume
independent of $Y$ given the covariates. The response could also be
interval-censored, meaning that it is only known to lie within an
interval $(Y^L,Y^U)$.

Such settings are not only common in survival analysis when studying
the time elapsed between a clearly defined time origin and an event of
interest, but also in surveys when the respondent reports a quantitive
response by pointing one interval or semi-interval in the partition of
the variable support.

We consider here a location-scale model,
\begin{align}
Y = \mu(\mathbf{z},\mathbf{x}) + \sigma(\mathbf{z},\mathbf{x})\varepsilon
\label{LocationScaleModel:Eq}
\end{align}
to describe the distribution of the response conditionally on the
covariates, where $\mu(\mathbf{z},\mathbf{x})$ denotes the conditional
location, $\sigma(\mathbf{z},\mathbf{x})$ the conditional dispersion,
and $\varepsilon$ an error term independent of $\mathbf{z}$ and
$\mathbf{x}$ assumed to have fixed 1st and 2nd order moments. One
could for example assume that $\E(\varepsilon)=0$ and
$\V(\varepsilon)=1$. The latter conditions lead to interpretation of
$\mu(\mathbf{z},\mathbf{x})$ and $\sigma(\mathbf{z},\mathbf{x})$ as
the conditional mean and standard deviation, respectively. Other
constraints are possible such as in \citet{Lambert2013} where
$\varepsilon$ was assumed to have a zero median and a unit
interquantile range, implying that $\mu(\mathbf{z},\mathbf{x})$ and
$\sigma(\mathbf{z},\mathbf{x})$ had to be interpreted as the
conditional median and interquantile range.

Assume that independent copies $(y_i,\mathbf{z}_i,\mathbf{x}_i)$
($i=1,\ldots,n$) are observed on $n$ units with the possibility of
right or interval censoring on $y_i$  as described above.
We consider additive models for the conditional location and
dispersion of the response:
\begin{align}
\big(\mu(\mathbf{z}_{i}, \mathbf{x}_i)\big)_{i=1}^n
= \left(\beta_{0}+\sum_{k=1}^p\beta_k z_{ik}
+\sum_{j=1}^{J} {f_{j}^{\mu}}(x_{ij})\right)_{i=1}^n
=\mathbf{Z} \pmb{\beta}
+\sum_{j=1}^{J}\mathbf{f}^\mu_j \label{CondLoc:Eq}\\
\big(\log{\sigma(\mathbf{z}_{i}, \mathbf{x}_i)}\big)_{i=1}^n
= \left(\delta_{0}+\sum_{k=1}^p \delta_k z_{ik}
+\sum_{j=1}^{J} {{f_{j}^{\sigma}}}(x_{ij})  \right)_{i=1}^n
= \mathbf{Z} \pmb{\delta}
+\sum_{j=1}^{J}\mathbf{f}^\sigma_j  \label{CondDisp:Eq}
\end{align}
where $f_{j}^{\mu}(\cdot)$ and $f_{j}^{\sigma}(\cdot)$ denote smooth
additive terms quantifying the effect of the $j$th quantitative
covariate on the conditional mean and dispersion,
$\mathbf{f}^\mu_j=\big(f_j^\mu(x_{ij})\big)_{i=1}^n$ and
$\mathbf{f}^\sigma_j=\big(f_j^\sigma(x_{ij})\big)_{i=1}^n$ their
values over units stacked in vectors, $\mathbf{Z}$ the $n\times (1+p)$ design matrix
with a column of 1's for the intercept and one column per additional categorical covariate.
For simplicity and without restriction, we assume that the
quantitative covariates take values in $(0,1)$. This can be achieved
for $x_j$ by relocating and rescaling it using
e.g.\ the following linear transform,
$(x_j-\min_i\{x_{ij}\})/(\max_i\{x_{ij}\}-\min_i\{x_{ij}\})$.
Now consider a basis of $(L+1)$ cubic B-splines
$\{s^*_\ell(\cdot)\}_{\ell=1}^{L+1}$ associated to equally spaced knots on
$(0,1)$. They are recentered for identification purposes in the additive model using
$s_\ell(\cdot)= s^*_{\ell}(\cdot) - \int_0^1s^*_{\ell}(u)du~(\ell=1,\ldots,L)$.
Then, the additive terms in the conditional location and
dispersion models can be approximated using linear combinations of these (recentered)
B-splines,
%\begin{align*}
$\mathbf{f}^\mu_j=\left(\sum_{\ell=1}^L s_\ell(x_{ij})\theta_{\ell
    j}^\mu\right)_{i=1}^n=\mathbf{S}_j\pmb{\theta}_j^\mu$, %~~;~~
$\mathbf{f}^\sigma_j=\left(\sum_{\ell=1}^L s_\ell(x_{ij})\theta_{\ell j}^\sigma\right)_{i=1}^n=\mathbf{S}_j\pmb{\theta}_j^\sigma,$
%\end{align*}
where $[\mathbf{S}_j]_{i\ell}=s_\ell(x_{ij})$,
$\big(\pmb{\theta}_j^\mu\big)_\ell=\theta_{\ell j}^\mu$
and $\big(\pmb{\theta}_j^\sigma\big)_\ell=\theta_{\ell j}^\sigma$.
Hence, using vectorial notations, the expressions for the conditional
location and dispersion in (\ref{CondLoc:Eq}) and (\ref{CondDisp:Eq})
can be rewritten as
%\begin{align*}
$\big(\mu_i=\mu(\mathbf{z}_{i}, \mathbf{x}_i)\big)_{i=1}^n
={\mathbfcal X} \pmb{\psi}^\mu$, %~~;~~
$\big(\sigma_i = \sigma(\mathbf{z}_{i}, \mathbf{x}_i)\big)_{i=1}^n
=\exp\big({\mathbfcal X} \pmb{\psi}^\sigma\big)$
%\end{align*}
with design matrix ${{\mathbfcal X}}=[\mathbf{Z},\mathbf{S}_1,\ldots,\mathbf{S}_J]
=[\mathbf{Z},\mathbfcal{S}]\in\rit^{n\times q}$;
matrices of spline parameters (with one column per additive term)
$\mathbf{\Theta}^\mu=[\pmb{\theta}_1^\mu,\ldots,\pmb{\theta}_J^\mu]$,
$\mathbf{\Theta}^\sigma=[\pmb{\theta}_1^\sigma,\ldots,\pmb{\theta}_J^\sigma]$ in $\rit^{L\times J}$;
vectors of (stacked) regression parameters
$\pmb{\psi}^\mu = \begin{pmatrix}\pmb{\beta}, \Vec{\mathbf{\Theta}^\mu}\end{pmatrix}$,
$\pmb{\psi}^\sigma = \begin{pmatrix}\pmb{\delta}, \Vec{\mathbf{\Theta}^\sigma}\end{pmatrix}$
in $\rit^q$, where $q=(1+p+JL)$.
With $p_1$ (resp.\,$p_2$) covariates and a B-spline basis of size $L_1$ (resp.\,$L_2$)
shared by each of the $J_1$ (resp.\,$J_2$) additive terms in the location (resp.\,dispersion) model,
we would end up with design matrices
${\mathbfcal X}^\mu=[\mathbf{Z}^\mu,\mathbfcal{S}^\mu]\in\rit^{n\times q_1}$
(resp.\,${\mathbfcal X}^\sigma=[\mathbf{Z}^\sigma,\mathbfcal{S}^\sigma]\in\rit^{n\times q_2}$) with
$q_1=(1+p_1+J_1L_1)$ (resp.\,$q_2=(1+p_2+J_2L_2)$) such that
%\begin{align*}
$\big(\mu_i\big)_{i=1}^n
={\mathbfcal X}^\mu \pmb{\psi}^\mu,$ % ~~;~~
$\big(\sigma_i\big)_{i=1}^n
=\exp\big({\mathbfcal X}^\sigma \pmb{\psi}^\sigma\big)
$.
%\end{align*}

%%
\subsection{Penalized log-likelihood for the joint regression model}

Estimation of the regression parameters and of the additive terms (for
given penalty parameters) can
be made using penalized likelihood.  Denote by
$f_\epsilon(\cdot~;\pmb{\phi})$
(resp.\,$S_\epsilon(\cdot~;\pmb{\phi})$)
the conditional density (resp.\,survival function) of the standardized error term $\epsilon$ in
(\ref{LocationScaleModel:Eq}) with a possible dependence on a set of
parameters $\pmb{\phi}$.
The contribution
$\ell_i=\ell_i(\pmb{\psi}^\mu,\pmb{\psi}^\sigma,\pmb{\phi};\D)$
of unit $i$ to the
log-likelihood will depend on the censoring status of the observed response $y_i$:
%\vspace{-\topsep}
\begin{description}\itemsep0em
\item[--] {Uncensored} $y_i=t_i$: then, the corresponding standardized
  error term $e_i$ is equal to $r_i=(y_i-\mu_i)/\sigma_i$ with log-likelihood contribution
$\ell_i=-\log\sigma_i+\log f_\epsilon(r_i)$.

\item[--] Right-censored at $y_i>t_i$: then, the corresponding standardized error term is
$e_i> r_i=(t_i-\mu_i)/\sigma_i$ with log-likelihood contribution
$\ell_i=\log S_\epsilon(r_i)$.

\item[--] Interval-censored with $y_i\in(y_{i}^{L},y_{i}^{R})$: then, the log-likelihood contribution is\,
$\ell_i=\log\left(S(r_i^{L})-S(r_i^{R})\right)$ as $e_i\in (r_i^L,r_i^R)$ where
$r_i^L=(y_i^L-\mu_i)/\sigma_i$ and $r_i^R=(y_i^R-\mu_i)/\sigma_i$.
\end{description}
Smoothness of the additive terms can be tuned by penalizing changes in
differences of neighbour spline parameters
\citep{Eilers1996,Eilers2010}. In a frequentist framework, this can be
done by adding one penalty (to the log-likelihood) per additive
term. When penalizing second-order differences in the location model,
the penalty for the $j$th additive term ($j=1,\ldots,J_1$) becomes
%\begin{align*}
$\lambda^{\mu}_{j}
\sum_{\ell=1}^{L_1-2}\{(\theta^\mu_{\ell+2,j}-\theta^\mu_{\ell+1,j})-(\theta^\mu_{\ell+1,j}-\theta^\mu_{\ell,j})\}^2
= \lambda^{\mu}_{j} \sum_{\ell}\left(\mathbf{D}^\mu\pmb{\theta}^\mu_{j}\right)_{\ell}^{2}
= {\pmb{\theta}^\mu_{j}}^{\top}
  (\lambda^{\mu}_{j}{\mathbf{P}^\mu})\pmb{\theta}^\mu_{j}
$, %\end{align*}
where $\mathbf{D}^\mu$ denotes the corresponding difference matrix and
$\mathbf{P}^\mu=(\mathbf{D}^\mu)^\top\mathbf{D}^\mu$ the associated penalty matrix.
At the limit, as $\lambda^{\mu}_{j}\rightarrow +\infty$, the estimated
second-order differences will tend to zero, forcing the estimate of
the function $f_{j}^{\mu}(x_{j})$ to be linear. Similar penalties with
penalty parameters $\lambda^\sigma_j$ can be defined for each additive
term in the dispersion model.

\subsection{Bayesian specification}
In a Bayesian framework, similar penalties arise through the specification of
conditional priors for the
spline parameters \citep{Lang2004}, yielding for the $j$th additive
terms in the location and dispersion models,
%\begin{align*}
$p(\pmb{\theta}^\mu_{j}|\lambda_j^{\mu}) \propto
% (\lambda^{\mu}_{j})^{\rho(\mathbf{P})/2}
\exp\left(-{1\over 2}~{\pmb{\theta}^\mu_{j}}^\top (\lambda^{\mu}_{j}\mathbf{P}^\mu) \pmb{\theta}^\mu_{j} \right),~
%~~;~~
p(\pmb{\theta}^\sigma_{j}|\lambda_j^{\sigma}) \propto
% (\lambda^{\mu}_{j})^{\rho(\mathbf{P})/2}
\exp\left(-{1\over 2}~{\pmb{\theta}^\sigma_{j}}^\top (\lambda^{\sigma}_{j}{\mathbf{P}^\sigma}) \pmb{\theta}^\sigma_{j} \right).
$ %\end{align*}
Assuming joint Normal priors for the intercepts and the regression
parameters associated to the other covariates $\mathbf{z}$,~
%$$
$\pmb{\beta}
\sim \N{\tilde{\mathbf{b}}}{{(\mathbf{Q}^\mu)}^{-1}},~
%~~;~~
\pmb{\delta}
\sim \N{\tilde{\mathbf{d}}}{{(\mathbf{Q}^{\sigma})}^{-1}},
$
%$$
the joint
priors for the regression and spline parameters in $\pmb{\psi}^\mu$ and
$\pmb{\psi}^\sigma$ induce Gaussian Markov random fields (GMRF)
\citep{RueHeld2005} as they can be written as
\begin{align*}
&
p(\pmb{\psi}^\mu|\pmb{\lambda}^{\mu}) \propto
\exp\left(-{1\over 2}~({\pmb{\psi}^\mu}-\mathbf{b})^\top \mathbf{K}^\mu_\lambda
({\pmb{\psi}^\mu}-\mathbf{b})
\right)~;~  \\
&
p(\pmb{\psi}^\sigma|\pmb{\lambda}^{\sigma}) \propto
\exp\left(-{1\over 2}~({\pmb{\psi}^\sigma}-\mathbf{d})^\top \mathbf{K}^\sigma_\lambda
({\pmb{\psi}^\sigma}-\mathbf{d})
\right),~
\end{align*}
where~
$\mathbf{b}=(\tilde{\mathbf{b}},\mathbf{0}_{J_1L_1})$,
$\mathbf{K}^\mu_\lambda= \diag\big(\mathbf{Q}^\mu,{\mathbfcal{P}}^\mu_\lambda\big)$,
${\mathbfcal{P}^\mu_\lambda}=\pmb{\Lambda}^\mu \kron \mathbf{P}^\mu$,
$[\pmb{\Lambda}^\mu]_{jj'}=\delta_{jj'}\lambda^\mu_j$,
$\mathbf{d}=(\tilde{\mathbf{d}},\mathbf{0}_{J_2L_2})$,
$\mathbf{K}^\sigma_\lambda= \diag\big(\mathbf{Q}^\sigma,{\mathbfcal{P}}^\sigma_\lambda\big)$,
${\mathbfcal{P}^\sigma_\lambda}=\pmb{\Lambda}^\sigma \kron \mathbf{P}^\sigma$
and $[\pmb{\Lambda}^\sigma]_{jj'}=\delta_{jj'}\lambda^\sigma_j$.
Then the joint posterior for the %whole set of
parameters %in the additive location-scale model
is
\begin{align*}
p(\pmb{\psi}^\mu,\pmb{\psi}^\sigma,\pmb{\lambda}^\mu,\pmb{\lambda}^\sigma,\pmb{\phi}|\D)
\propto
L(\pmb{\psi}^\mu,\pmb{\psi}^\sigma,\pmb{\phi};\D)
\,p(\pmb{\psi}^\mu|\pmb{\lambda}^\mu)
\,~p(\pmb{\psi}^\sigma|\pmb{\lambda}^\sigma)
\,~p(\pmb{\lambda}^\mu)\,p(\pmb{\lambda}^\sigma)\,p(\pmb{\phi}).
% ~p(\pmb{\phi})
%\label{JointPosterior:Eq}
\end{align*}

\subsection{Estimation of  $\pmb{\psi}^\mu$  and  $\pmb{\psi}^\sigma$}
\label{PsiMuPsiSigmaEstimation:Sec}

The estimation of the regression parameter $\pmb{\psi}^\mu$  and
$\pmb{\psi}^\sigma$ will be made sequentially and conditionally on the
error density $f_\epsilon(\cdot~;\pmb{\phi})$ and the penalty
parameters $\pmb{\lambda}^\mu$ and $\pmb{\lambda}^\sigma$.
It is based on the following decomposition of their joint conditional
posterior:
\begin{align}
p(\pmb{\psi}^\mu,\pmb{\psi}^\sigma|\pmb{\lambda}^\mu,\pmb{\lambda}^\sigma,\pmb{\phi},\D)
&=
p(\pmb{\psi}^\mu|\pmb{\psi}^\sigma,\pmb{\lambda}^\mu,\pmb{\phi},\D)
\,p(\pmb{\psi}^\sigma|\pmb{\lambda}^\mu,\pmb{\lambda}^\sigma,\pmb{\phi},\D).
\label{ConditionalPsiMuPsiSigma:Eq}
\end{align}
The conditional posterior for the location parameters is given by
\begin{align}
\label{CondPostPsiMu:Eq}
&p(\pmb{\psi}^\mu|\pmb{\psi}^\sigma,\pmb{\lambda}^\mu,\pmb{\phi},\D)
  \propto L(\pmb{\psi}^\mu,\pmb{\psi}^\sigma,\pmb{\phi};\D)
   ~p(\pmb{\psi}^\mu|\pmb{\lambda}^\mu),
\end{align}
while the final expression for
\begin{align}
p(\pmb{\psi}^\sigma|\pmb{\lambda}^\mu,\pmb{\lambda}^\sigma,\pmb{\phi},\D)
&=
{p(\pmb{\psi}^\mu,\pmb{\psi}^\sigma|\pmb{\lambda}^\mu,\pmb{\lambda}^\sigma,\pmb{\phi},\D)
  \over
 p(\pmb{\psi}^\mu|\pmb{\psi}^\sigma,\pmb{\lambda}^\mu,\pmb{\phi},\D)
}
\propto L(\tilde{\pmb{\psi}}^\mu,\pmb{\psi}^\sigma,\pmb{\phi};\D)
  \,p(\pmb{\psi}^\sigma|\pmb{\lambda}^\sigma)
  \begin{vmatrix} \tilde{\Sigma}^\mu_\lambda \end{vmatrix}^{1/2}
\label{PsiSigmaMarginalApprox2:Eq}
\end{align}
is obtained by using the Laplace approximation
$\N{\tilde{\pmb{\psi}^\mu_\lambda}}{\tilde{\Sigma}^\mu_\lambda}$
in the denominator %of (\ref{PsiSigmaMarginalApprox1:Eq})
and evaluating it at the posterior
mode $\tilde{\pmb{\psi}^\mu_\lambda}$.  Indeed, given the Normality
assumption for the prior $(\pmb{\psi}^\sigma|\pmb{\lambda}^\mu)$, the
conditional posterior in the denominator will be approximately Normal,
see \citet{RueMartino2009} for arguments in the general context of
Gaussian random fields.

Estimates for the regression parameters will be obtained by
alternating the maximization of (\ref{CondPostPsiMu:Eq}) and
(\ref{PsiSigmaMarginalApprox2:Eq}) till convergence.
For $\pmb{\psi}^\mu$, this is done for given values of
the other parameters using a Newton-Raphson (N-R) algorithm built upon
the gradient and (minus) Hessian of the log
of (\ref{CondPostPsiMu:Eq}),
\begin{align} \label{GradHesPsiMu:Eq}
\begin{split}
% Grad psi^mu
{\mathbf{U}}^\lambda_{{\psi}^\mu}(\pmb{\psi}^\mu)&=
{\partial \log p(\pmb{\psi}^\mu|\pmb{\lambda}^\mu,\pmb{\psi}^\sigma,\pmb{\phi},\D)
 \over \partial \pmb{\psi}^\mu}
= {{\mathbfcal X}^\mu}^\top \pmb{\omega}^\mu -\mathbf{K}^\mu_\lambda\pmb{\psi}^\mu, \\
% Hes psi^mu
-{\mathbf{H}}^\lambda_{{\psi}^\mu}(\pmb{\psi}^\mu)&
= -{\partial^2 \log p(\pmb{\psi}^\mu|\pmb{\lambda}^\mu,\pmb{\psi}^\sigma,\pmb{\phi},\D)
 \over \partial \pmb{\psi}^\mu \partial {\pmb{\psi}^\mu}^\top}
={{\mathbfcal X}^\mu}^\top \mathbf{W}^\mu{\mathbfcal X}^\mu+\mathbf{K}^\mu_\lambda~,
\end{split}
\end{align}
with $\pmb{\omega}^\mu\in\rit^n$ and $\mathbf{W}^\mu=\diag(\mathbf{w}^\mu)\in\rit^{n\times n}$
given in Appendix \ref{Appendix:A}. At convergence, it yields the conditional posterior
mode $\tilde{\pmb{\psi}}^\mu$ and variance-covariance matrix
$\tilde{\Sigma}^\mu_\lambda= \left(-\mathbf{H}^\lambda_{\psi^\mu}(\tilde{\pmb{\psi}}^\mu)\right)^{-1}$
in the above mentioned Laplace approximation.
The estimates for $\pmb{\psi}^\sigma$ are also obtained using a N-R
algorithm based on the gradient and (minus) Hessian of the log of (\ref{PsiSigmaMarginalApprox2:Eq}),
\begin{align} \label{GradHesPsiSigma:Eq}
%\begin{split}
% Grad psi^sigma
{\mathbf{U}}^\lambda_{{\psi}^\sigma}(\pmb{\psi}^\sigma)&=
{\partial \log p(\pmb{\psi}^\sigma|\pmb{\lambda}^\sigma,\pmb{\psi}^\mu,\pmb{\phi},\D)
 \over \partial \pmb{\psi}^\sigma}
= {{\mathbfcal X}^\sigma}^\top \pmb{\omega}^\sigma
-\mathbf{K}^\sigma_\lambda\pmb{\psi}^\sigma
+ {\partial E^\mu_\lambda\over \partial \pmb{\psi}^\sigma}
,\\
% Hes psi^sigma
-{\mathbf{H}}^\lambda_{{\psi}^\sigma}(\pmb{\psi}^\sigma)&
= -{\partial^2 \log p(\pmb{\psi}^\sigma|\pmb{\lambda}^\sigma,\pmb{\psi}^\mu,\pmb{\phi},\D)
 \over \partial \pmb{\psi}^\sigma \partial {\pmb{\psi}^\sigma}^\top}
={{\mathbfcal X}^\sigma}^\top \mathbf{W}^\sigma{\mathbfcal X}^\sigma+\mathbf{K}^\sigma_\lambda~
-{\partial^2 E^\mu_\lambda\over \partial \pmb{\psi}^\sigma \partial{\pmb{\psi}^\sigma}^\top}
\nonumber
%\end{split}
\end{align}
with $E^\mu_\lambda={1\over 2}\log\begin{vmatrix} \tilde{\Sigma}^\mu_\lambda \end{vmatrix}$
(see Appendix \ref{Appendix:B} for its partial derivatives) and
$\pmb{\omega}^\sigma\in\rit^n$, $\mathbf{W}^\sigma=\diag(\mathbf{w}^\sigma)\in\rit^{n\times n}$
defined in Appendix \ref{Appendix:A}.
It leads to Algorithm \ref{PsiMuPsiSigma:Algo}
for the estimation of the regression parameters
$\pmb{\psi}^\mu$ and $\pmb{\psi}^\sigma$.
\begin{algo} \label{PsiMuPsiSigma:Algo}
{\it Estimation of $\pmb{\psi}^\mu$ and $\pmb{\psi}^\sigma$}  \\
\textrm
At each iteration of their respective Newton-Raphson algorithm (and conditionally on
the values of the other model parameters):
\vspace{-\topsep}
\begin{enumerate} \itemsep0em
\item Compute the standardized residuals
     $r_i=(y_i-\mu_i(\pmb{\psi}^\mu_\lambda))/\sigma_i(\pmb{\psi}^\sigma_\lambda)$
     accompanied by their observation indicators $d_i$, some of these residuals
     being right-censored ($d_i=0$) or interval-censored with $r_i\in (r_i^L,r_i^R)$.

\item Recompute vectors $\pmb{\omega}^\mu$, $\pmb{\omega}^\sigma$,
  $\mathbf{w}^\mu$, $\mathbf{w}^\sigma$ and, hence, the diagonal
  matrices $\mathbf{W}^\mu=\diag(\mathbf{w}^\mu)$ and
  $\mathbf{W}^\sigma=\diag(\mathbf{w}^\sigma)$ using
  (\ref{OmegaW:UncensRightCens:eq}) and (\ref{OmegaW:IntCens:eq}).

\item Update the location and dispersion parameters $\pmb{\psi}^\mu_\lambda$
  and $\pmb{\psi}^\sigma_\lambda$ using
  (\ref{GradHesPsiMu:Eq}) and (\ref{GradHesPsiSigma:Eq}), % as follows,
 $
 \pmb{\psi}^\mu_\lambda \longleftarrow \pmb{\psi}^\mu_\lambda -
\big({\mathbf{H}}^\lambda_{{\psi}^\mu}\big)^{-1}{\mathbf{U}}^\lambda_{{\psi}^\mu}
~~;~~
 \pmb{\psi}^\sigma_\lambda \longleftarrow \pmb{\psi}^\sigma_\lambda -
\big({\mathbf{H}}^\lambda_{{\psi}^\sigma}\big)^{-1}{\mathbf{U}}^\lambda_{{\psi}^\sigma}~,
$
with step-halving when found necessary through the monitoring of
$p(\pmb{\psi}^\mu|\pmb{\psi}^\sigma,\pmb{\lambda}^\mu,\pmb{\phi},D)$ and
$p(\pmb{\psi}^\sigma|\pmb{\lambda}^\mu,\pmb{\lambda}^\sigma,\pmb{\phi},D)$,
respectively.
\end{enumerate}
At convergence, after a few iterations, one obtains
the conditional posterior modes $\hat{\pmb{\psi}}^\mu_\lambda$ and
$\hat{\pmb{\psi}}^\sigma_\lambda$ with negative inverse Hessians
$\Sigma^\mu_\lambda=\big(-{\mathbf{H}}^\lambda_{{\psi}^\mu}(\hat{\pmb{\psi}}^\mu_\lambda)\big)^{-1}$
and
$\Sigma^\sigma_\lambda=\big(-{\mathbf{H}}^\lambda_{{\psi}^\sigma}(\hat{\pmb{\psi}}^\sigma_\lambda)\big)^{-1}$.
%$\square$
\end{algo}

\subsection{Selection of the penalty parameters   $\pmb{\lambda}^\mu$  and  $\pmb{\lambda}^\sigma$}
\label{MarginalPostLambda:Sec}
Let $\pmb{\psi}=(\pmb{\psi}^\mu,\pmb{\psi}^\sigma)$ and
$\pmb{\lambda}=(\pmb{\lambda}^\mu,\pmb{\lambda}^\sigma)$.  Starting
from the joint posterior for the model parameters, we have (with an
implicit dependence on the standardized error distribution and its
parameter(s) $\pmb{\phi}$) the following identity for the marginal
posterior of $\pmb{\lambda}$: %the penalty parameters:
\begin{align}
p(\pmb{\lambda}|\D) =
{p(\pmb{\psi},\pmb{\lambda}|\D)
\over
p(\pmb{\psi}|\pmb{\lambda},\D)
}. \label{MarginalLambda:Eq1}
\end{align}
Given the conditional GMRF prior for $\pmb{\psi}$, we conclude that the
conditional posterior in the denominator is approximately Gaussian \citep{RueMartino2009}.
Using a Laplace approximation, we obtain
% \begin{align*}
$(\pmb{\psi} | \pmb{\lambda},\D)
\mathrel{\dot\sim}
\N{\hat{\pmb{\psi}}_\lambda}{\Sigma_\lambda},$
% \end{align*}
where $\hat{\pmb{\psi}}_\lambda$ denotes the conditional posterior mode of
$\pmb{\psi}$ (obtained using Algorithm \ref{PsiMuPsiSigma:Algo})
and
$$\Sigma_\lambda^{-1} = -\sum_{i=1}^n
\begin{bmatrix}
{\partial^2 \ell_i\over \partial\psi^\mu \partial(\psi^\mu)^\top}
& {\partial^2 \ell_i\over \partial\psi^\mu \partial(\psi^\sigma)^\top} \\
{\partial^2 \ell_i\over \partial\psi^\mu \partial(\psi^\sigma)^\top}
& {\partial^2 \ell_i\over \partial\psi^\sigma \partial(\psi^\sigma)^\top}
\end{bmatrix}
+
\begin{bmatrix}
\mathbf{K}^\mu_\lambda & \cdot\\
\cdot & \mathbf{K}^\sigma_\lambda
\end{bmatrix},
$$
see also \citet{Tierney:Kadane:1986} for general arguments for such an
approximation to the marginal posterior of $\pmb{\lambda}$.
Evaluating the RHS of (\ref{MarginalLambda:Eq1}) at $\hat{\pmb{\psi}}_\lambda$
with the preceding Laplace approximation,
we approximate $p(\pmb{\lambda}|\D)$ by
%we obtain the following approximation to $p(\pmb{\lambda}|\D)$:
%\begin{align}
$
\tilde{p}(\pmb{\lambda}|\D) \propto
p(\hat{\pmb{\psi}}_\lambda,\pmb{\lambda}|\D)
~{\begin{vmatrix}\Sigma_\lambda^{-1}\end{vmatrix}}^{-1/2}.
$
%\label{ApproxMarginalLambda:Eq1}
%\end{align}
\citet[][Section4]{Wood:Fasiolo:2017} obtained a similar starting
expression to build their proposal for the selection of penalty
parameters in an additive regression model with a parametric error
distribution.  \citet{Gressani:Lambert:2018} also followed that
strategy in the context of a cure survival model with splines used to
specify the baseline hazard function for susceptible subjects.
Ignoring the cross-derivatives
in
%of the log-likelihood in the expression for
$\Sigma_\lambda^{-1}$ yields
\begin{align}
\tilde{p}(\pmb{\lambda}|\D) \propto
p(\hat{\pmb{\psi}}_\lambda,\pmb{\lambda}|\D)
~{\begin{vmatrix}{{\mathbfcal X}^\mu}^\top \mathbf{W}^\mu{\mathbfcal X}^\mu+\mathbf{K}^\mu_\lambda\end{vmatrix}}^{-1/2}
~{\begin{vmatrix}{{\mathbfcal X}^\sigma}^\top \mathbf{W}^\sigma{\mathbfcal X}^\sigma+\mathbf{K}^\sigma_\lambda\end{vmatrix}}^{-1/2}
, \label{ApproxMarginalLambda:Eq2}
\end{align}
with
%expressions for
$\mathbf{W}^\mu$ and $\mathbf{W}^\sigma$
given in Appendix \ref{Appendix:A}.
Dropping the $\mu$ or $\sigma$ superscript and letting
\begin{align} \label{Mcal:Eq}
&\mathbfcal{M} =
 {\mathbfcal S}^\top \mathbf{W}{\mathbfcal S}
-{\mathbfcal S}^\top \mathbf{W}{\mathbf Z}
   ({\mathbf Z}^\top\mathbf{W}{\mathbf Z}+\mathbf{Q})^{-1}
  {\mathbf{Z}}^\top\mathbf{W}{\mathbfcal S},
\end{align}
each determinant in (\ref{ApproxMarginalLambda:Eq2}) can be rewritten as~
\begin{align*}
{\begin{vmatrix}{\mathbfcal X}^\top \mathbf{W}{\mathbfcal X}+\mathbf{K}_\lambda\end{vmatrix}}
=
{\begin{vmatrix}{{\mathbf Z}}^\top \mathbf{W}{\mathbf Z} + \mathbf{Q}\end{vmatrix}}
~
{\begin{vmatrix}{{\mathbfcal M}} +  \mathbfcal{P}_\lambda\end{vmatrix}},
\end{align*}
where only the last factor directly depends on the penalty parameters $\pmb{\lambda}$.
Combined with (\ref{ApproxMarginalLambda:Eq2}) and taking
$\lambda^\mu_j\sim\G{1}{b^\mu=10^{-4}}$, we conclude that
\begin{align}
&\log\tilde{p}(\pmb{\lambda}^\mu|\pmb{\lambda}^\sigma,\D)
\mathrel{\dot=}
\log p(\hat{\pmb{\psi}}_\lambda,\pmb{\lambda}|\D)
-{1\over 2}\log \begin{vmatrix}{{\mathbfcal M}^\mu} + \mathbfcal{P}^\mu_\lambda\end{vmatrix}
\label{MarginalLambdaMu:Eq}\\
~~&= \ell(\hat{\pmb{\psi}}_\lambda;\D)
+\sum_{j=1}^{J_1}\left\{
  {L_1-r\over 2}\log \lambda^\mu_j
-\left(b^\mu+{{1\over 2}({\hat{\pmb{\theta}}^\mu_{j\lambda}})^{\top}{\mathbf{P}^\mu}\hat{\pmb{\theta}}^\mu_{j\lambda}}\right) \lambda^{\mu}_{j}
  \right\} \nonumber\\
~~&~~
-{1\over 2}\log \begin{vmatrix}{{\mathbfcal M}^\mu} +  \mathbfcal{P}^\mu_\lambda\end{vmatrix}.
\nonumber
\end{align}
The indirect dependence of the log-likelihood and of ${\mathbfcal M}^\mu$ on
$\lambda^\mu$ (through $\hat{\pmb{\psi}}_\lambda$ and $\mathbf{W}^\mu$) will be ignored
during the computation of the gradient $\mathbf{U}_{{\lambda}^\mu}$ and Hessian
$\mathbf{H}_{{\lambda}^\mu}$ as (non reported) numerical simulations suggest that
this dependence is moderate. Practically, in an iterative
maximization of (\ref{MarginalLambdaMu:Eq}) using the N-R algorithm, we fix
$\ell(\hat{\pmb{\psi}}_\lambda;\D)$ and ${\mathbfcal M}^\mu$ at their
values $\breve{\ell}$ and
$\breve{\mathbfcal M}^\mu$ at the beginning of the iteration,
and compute the gradient and Hessian of
\begin{align*}
\log\breve{p}(\pmb{\lambda}^\mu|\pmb{\lambda}^\sigma,\D)%\\
=&~ \breve{\ell}
+\sum_{j=1}^{J_1}\left\{
  {L_1-r\over 2}\log \lambda^\mu_j
-\left(b^\mu+{{1\over 2}{({\hat{\pmb{\theta}}^\mu}_{j\lambda}})^{\top}{\mathbf{P}^\mu}{\hat{\pmb{\theta}}^\mu}_{j\lambda}}\right) \lambda^{\mu}_{j}
  \right\}\\
&-{1\over 2}\log \begin{vmatrix}{\breve{\mathbfcal M}^\mu} +  \mathbfcal{P}^\mu_\lambda\end{vmatrix}~.
\end{align*}
Let
$\breve{\cal R}_j^\mu = \breve{\cal R}_j^\mu(\pmb{\lambda}^\mu)=
\begin{pmatrix}{\breve{\mathbfcal M}^\mu} + \mathbfcal{P}^\mu_\lambda\end{pmatrix}^{-1}
\left((\pmb{1}_j\pmb{1}_j^\top)\kron \mathbf{P}^\mu\right)
$
for $j=1,\ldots,J_1$ where $\pmb{1}_j$ denotes the $j$th unit vector.
Then, using results on the derivative of determinants and after some
algebra, on can show that
\begin{align}
\big(\breve{\mathbf{U}}_{{\lambda}^\mu}(\pmb{\lambda}^\mu)\big)_j&=
{\partial \log\breve{p}(\pmb{\lambda}^\mu|\pmb{\lambda}^\sigma,\D)
 \over \partial {\lambda}^\mu_j}
= {L_1-r\over 2 \lambda^\mu_j}
-\left(b^\mu+{1\over 2}({\hat{\pmb{\theta}}^\mu_{j\lambda}})^{\top}{\mathbf{P}^\mu}\hat{\pmb{\theta}}^\mu_{j\lambda}\right)
-{1\over 2}\tr{\breve{\cal R}_j^\mu}, \nonumber\\
-[\breve{\mathbf{H}}_{{\lambda}^\mu}(\pmb{\lambda}^\mu)]_{jk}&
= -{\partial^2 \log\breve{p}(\pmb{\lambda}^\mu|\pmb{\lambda}^\sigma,\D)
 \over \partial {\lambda}^\mu_j \partial {\lambda}^\mu_k}
={L_1-r\over 2 {(\lambda^\mu_j)}^2}{\delta_{jk}}
-{1\over 2}\tr{\breve{\cal R}_j^\mu\breve{\cal R}_k^\mu}.
\label{GradHessianLogLambdaMu:Eq}
\end{align}
Similar expressions can be obtained for
$(\pmb{\lambda}^\sigma|\pmb{\lambda}^\mu,\D)$ by switching the role of
$\mu$ and $\sigma$ as superscripts. The penalty parameters are
selected to maximize (\ref{MarginalLambdaMu:Eq}) and its counterpart for
$\pmb{\lambda}^\sigma$ using Algorithm \ref{LambdaMuLambdaSigma:Algo}
, yielding $\hat{\pmb{\lambda}}^\sigma$ and $\hat{\pmb{\lambda}}^\sigma$.

\begin{algo} \label{LambdaMuLambdaSigma:Algo}
{\it Selection of $\pmb{\lambda}^\mu$ and $\pmb{\lambda}^\sigma$} \\
Let
$g(\pmb{\nu})=\log\tilde{p}(\pmb{\lambda}^\mu|\pmb{\lambda}^\sigma,\D)$
where $\pmb{\lambda}^\mu=\lambda_{\min}+\exp(\pmb{\nu})$ with
$\lambda_{\min}$ denoting the smallest desirable value for the
penalty parameter of an additive term.
Using the chain rule, one can show that
$
({\breve{\mathbf{U}}}_{\nu})_j %= {\partial \breve{g}(\pmb{\nu}) \over \partial {\lambda_j}}
=\exp(\nu_j) (\breve{\mathbf{U}}_{{\lambda}^\mu})_j
$
and
$
(\breve{\mathbf{H}}_{\nu})_{jk} %={\partial^2 \breve{g}(\pmb{\nu}) \over \partial {\lambda_j}\partial {\lambda_k}}
=\exp(\nu_j+\nu_k) (\breve{\mathbf{H}}_{{\lambda}^\mu})_{jk}+\delta_{jk}\exp(\nu_j)(\breve{\mathbf{U}}_{\nu})_j
$
for $1\leq j,k \leq J_1$.
We propose to select $\pmb{\lambda}^\mu$ by maximizing $g(\pmb{\nu})$
using a Newton-Raphson algorithm with at each iteration:
%\vspace{-\topsep}
\begin{enumerate} \itemsep0em
\item \vspace{-\topsep}
  \begin{enumerate} \itemsep0em
  \item Given current values for $\pmb{\lambda}^\mu$ and
    $\hat{\pmb{\theta}}^\mu_{\lambda}$, compute the gradient $\breve{\mathbf{U}}_{\nu}$
     and Hessian matrix $\breve{\mathbf{H}}_{\nu}$
    using (\ref{GradHessianLogLambdaMu:Eq}) ;
  \item Update:
    $
    {\pmb{\nu}} \longleftarrow
    {\pmb{\nu}}
    -\breve{\mathbf{H}}_{\nu}^{-1} \breve{\mathbf{U}}_{\nu}~;
    ~\pmb{\lambda}^\mu \longleftarrow \lambda_{\min}+\exp(\pmb{\nu})~;$
  \end{enumerate}
\item Update $\hat{\pmb{\psi}}^\mu_\lambda$ using
  Algorithm \ref{PsiMuPsiSigma:Algo},
 $\breve{\cal M}^\mu$ using (\ref{Mcal:Eq}), yielding
  $\hat{\pmb{\theta}}^\mu_{j\lambda}$ and $\breve{\cal R}_j^\mu$,
\end{enumerate}
giving at convergence
$\hat{\pmb{\lambda}}^\mu =\exp(\hat{\pmb{\nu}})$.\\
The same procedure with the superscripts $\sigma$ and $\mu$
interchanged yields $\hat{\pmb{\lambda}}^\sigma$.
\end{algo}

\subsection{Nonparametric pivotal density} \label{NPDensityEstimation:Sec}
\subsubsection{Density specification}
Besides classical parametric choices for the distribution of the
standardized error term $\epsilon$, nonparametric forms could be
preferred. Here, we propose to specify that distribution through the
associated hazard $h_\epsilon(\cdot)$ function using a linear
combination of $K$ B-splines,
$\log h_\epsilon(r) = \sum_{k=1}^K b_k(r) \phi_k$, where
$\{b_k(\cdot):k=1,\ldots,K\}$ denotes a large B-spline basis
associated to an equidistant grid of knots on the support of the
distribution. Given the constraints $\E(\varepsilon)=0$ and
$\V(\varepsilon)=1$, one can practically assume (using Chebyshev's
theorem) that (most of) the probability mass is on
$(r_{\min},r_{\max})=(-6,6)$, say.  Our approach is to some extent
connected to the proposal made by \citet{Cai:Hyndman:Wand:2002} with a
(truncated) linear spline basis in a mixed model framework. We go
further here by considering interval-censored data and moment
constraints for the underlying density function.  Note that starting
from the hazard function to estimate the underlying distribution does
not imply that the underlying variable must be positive. The only
requirement is the designation of a (conservative) lower bound for the
support of the standardized error term.
A spline approximation to the log-density could also be
considered \citep{Eilers1996, Kooperberg:Stone:1991,
  Lambert:Eilers:2009, Lambert:2011}, but a construct based on the
hazard function turns out to be analytically more convenient to handle
censored data, see below.

\subsubsection{Density estimation from i.i.d. right-censored data} \label{DensityEstRCData:Sec}
We now detail how we propose to estimate the spline coefficients $\pmb{\phi}$ in
the framework of Bayesian P-splines from potentially right- or even
interval-censored data.

Denote by $\{{\cal J}_j=[a_{j-1},a_j)\}_{j=1}^J$ a partition of
$(r_{\min} ,r_{\max})$ into a very large number $J$ of bins of equal
width $\Delta$ with midpoints $\{u_j\}_{j=1}^J$. Given a random sample
of $n$ i.i.d.\,observations $r_i$ ($i=1,\ldots,n$) for a potentially right-censored
(coded by $d_i=0$ and $1$ otherwise) variable $\varepsilon$,
%then, only known to be in $(r_i^L,r_i^R)$,
let $k_j=\sum_{i=1}^n k_{ij}$ and $n_j=\sum_{i=1}^n n_{ij}$
 with $k_{ij}=\mathbbm{1}(r_i\in{\cal J}_j)\,\mathbbm{1}(d_i=1)$
 and $n_{ij}=\mathbbm{1}(r_i\geq a_{j-1})=\mathbbm{1}(r_i\in \cup_{s\geq j}{\cal J}_s)$.
The log-likelihood for the estimation of the spline parameters
$\pmb{\phi}=(\phi_1,\ldots,\phi_K)$ from right-censored data can be written as
\begin{align}
\ell(\pmb{\phi}|\D) &= \sum_{i=1}^n \big\{d_i\log h_\epsilon(r_i) - H_\epsilon(r_i)\big\}
\approx \sum_{j=1}^J (k_j\log h_j - n_jh_j\Delta)
\label{DensityLoglik:Eq}
\end{align}
with $h_j=h_\epsilon(u_j)=\exp\{ \sum_{k=1}^K b_k(u_j) \phi_k\}$
where the approximation in (\ref{DensityLoglik:Eq}) comes from data
binning and quadrature to approximate the cumulated hazard function.
Following \citet{Eilers1996}, we penalize third order ($r=3$) differences of
successive spline parameters, yielding the penalized log-likelihood,
%\begin{align}
%&
$\ell_p(\pmb{\phi}|\tau,\D) = \ell(\pmb{\phi}|\D)
-{\tau\over 2} \pmb{\phi}^\top \mathbf{P} \pmb{\phi},
% \label{lpenDensityEst:Eq}
% \end{align}
$
with penalty matrix $\mathbf{P}$ of rank $(K-r)$.
Given the expressions for the gradient and Hessian,
\begin{align}
\mathbf{U}_{\tau}(\pmb{\phi})
%&=\nabla_\phi \ell_p
&={\partial \ell_p\over \partial \pmb{\phi}} = \mathbf{B}^\top(\mathbf{k}-\mathbf{n}\mathbf{h}\Delta) - \tau \mathbf{P}\pmb{\phi}
~;~\label{GradLpen:Eq}\\
-\mathbf{H}_{\tau}(\pmb{\phi})
%&=-\nabla_\phi^2 \ell_p
&=-{\partial^2 \ell_p\over \partial \pmb{\phi}\partial \pmb{\phi}^\top}
= \mathbf{B}^\top \diag(\mathbf{n}\mathbf{h}\Delta) \mathbf{B} + \tau\mathbf{P},
\label{HesLpen:Eq}
\end{align}
where $[\mathbf{B}]_{jk}=b_k(u_j)$, $\mathbf{k}=(k_j)_{j=1}^J$,
$\mathbf{n}=(n_j)_{j=1}^J$, $\mathbf{h}=(h_j)_{j=1}^J$,
one can use the (fast converging) Newton-Raphson procedure to obtain
spline parameter estimates for a given value of the penalty parameter
$\tau$, with at each iteration,
$\pmb{\phi} \longleftarrow \pmb{\phi} -\big(\mathbf{H}_\tau(\pmb{\phi})\big)^{-1}\mathbf{U}_\tau(\pmb{\phi})$,
yielding at convergence $\hat{\pmb{\phi}}_\tau$.

\subsubsection{Inclusion of interval-censored data} \label{DensityEstICData:Sec}
The contribution of interval-censored units to $k_j$ and $n_j$ can
also be included and reevaluated at every iteration of the preceding
Newton-Raphson procedure. Denote the hazard and density estimates
from the previous iteration by $\tilde{h}_\epsilon(\cdot)$ and
$\tilde{f}_\epsilon(\cdot)=\tilde{h}_\epsilon(\cdot)\exp(-\tilde{H}_\epsilon(\cdot))$,
 and let
$\tilde\pi_j=\int_{{\cal J}_j} \tilde{f}_\epsilon(r)dr \approx
\tilde{f}_\epsilon(u_j)\Delta$.
Consider an interval-censored observation $r_i\in(r_i^L,r_i^R)$ and
let ${\cal G}_i=\{j: {\cal J}_j\cap (r_i^L,r_i^R)\neq \emptyset \}$.
Then, the contribution of unit $i$ to the previously defined $k_j$ and
$n_j$ are given by
$k_{ij} = {\tilde\pi_j / \sum_{s\in{\cal G}_i}\tilde\pi_s}~
\mathbbm{1}(j \in {\cal G}_i)$
and
$n_{ij} = \mathbbm{1}(j<\min{\cal G}_i)
 + {\sum_{s=j}^{\max{\cal G}_i}\tilde\pi_s / \sum_{s\in{\cal G}_i}\tilde\pi_s}
   \mathbbm{1}(j\in{\cal G}_i)$, repectively.
At convergence, the procedure in Section \ref{DensityEstRCData:Sec}
with, now, interval-censored data entering the computation of $k_j$
and $n_j$ will provide an estimate $\hat{\pmb{\phi}}_\tau$ of the spline
parameters $\pmb{\phi}$ for given $\tau$ and, hence, of the density estimate
underlying the potentially right- or interval-censored observations.

\subsubsection{Density estimation with moment constraints}
Constraints on the mean and variance of the underlying distribution can also be
forced. More generally, consider a set of (potentially) nonlinear constraints
$F_s(\pmb{\phi})=f_s~(s=1,\ldots,S)$ shortly denoted vectorially by
$\pmb{F}(\pmb{\phi})=\mathbf{f}$. At every iteration of the preceding
Newton-Raphson procedure, we suggest to linearize each constraint
using a first-order Taylor expansion about the current estimate
$\tilde{\pmb{\phi}}$ of the spline parameters,
%$$
$
\tilde{F}_s(\pmb{\phi}) = F_s(\tilde{\pmb{\phi}})
 +\tilde{\mathbf{v}}_s^\top(\pmb{\phi}-\tilde{\pmb{\phi}})
~~\text{with}~~\tilde{\mathbf{v}}_s
={\partial F_s(\tilde{\pmb{\phi}})\over \partial \pmb{\phi}}.
$
%$$
Hence, letting $\tilde{\mathbf V}
  =[\tilde{\mathbf{v}}_1,\ldots,\tilde{\mathbf{v}}_S]^\top
  \in \rit^{S\times K}$,
a linearized version of the constraints is
%$$
$\tilde{\mathbf V}\pmb{\phi}=\tilde{\mathbf{c}}
~~\text{with}~~
\tilde{\mathbf{c}}=\tilde{\mathbf{V}} \tilde{\pmb{\phi}}
   + (\mathbf{f}-\pmb{F}(\tilde{\pmb{\phi}})).
$
%$$
The estimation of the spline parameters under these linearized
constraints can be made using the Lagrangian
\begin{align}
G(\pmb{\phi},\pmb{\omega}) = \ell_p(\pmb{\phi}|\tau,\D)
 -\pmb{\omega}^\top(\tilde{\mathbf V}\pmb{\phi}-\tilde{\mathbf{c}}),
\label{DensityEstimation:Lagrangian:Eq}
\end{align}
with Lagrange multipliers $\pmb{\omega}$. Practically, at every
iteration of a Newton-Raphson procedure, the preceding values
$(\tilde{\pmb{\phi}},\tilde{\pmb{\omega}})$ of the spline parameters
and Lagrange multipliers are updated using
\begin{align}
\begin{pmatrix} \tilde{\pmb{\phi}} \\ \tilde{\pmb{\omega}} \end{pmatrix}
\longleftarrow
\begin{pmatrix} \tilde{\pmb{\phi}} \\ \tilde{\pmb{\omega}} \end{pmatrix}
-
\begin{pmatrix}
{\partial^2 \ell_p(\tilde{\pmb{\phi}}|\tau,\D)\over \partial \pmb{\phi}\partial \pmb{\phi}^\top} & -\tilde{\mathbf V}^\top
\\[1em] % Add some extra space between lines
-\tilde{\mathbf V} & \mathbf{0}
\end{pmatrix}^{-1}
\begin{pmatrix}
{\partial \ell_p(\tilde{\pmb{\phi}}|\tau,\D)\over \partial \pmb{\phi}}-\tilde{\mathbf V}^\top\tilde{\pmb{\omega}}
\\[1em] % Add some extra space between lines
-\tilde{\mathbf V}\tilde{\pmb{\phi}}+\tilde{\mathbf{c}}
\end{pmatrix},
\label{ConstrainedDensityEst:Eq}
\end{align}
with partial derivatives of the penalized log-likelihood given in
(\ref{GradLpen:Eq}) and (\ref{HesLpen:Eq}).

Now consider specific constraints on the spline parameters based on
the first two moments ($S=2$) of the density, remembering that
$f(u_j)=h_j\exp(-H_j)$ (and letting $\Delta \rightarrow 0^+$):
\begin{align*}
&
\E(\epsilon)=\mu_\epsilon=0
\Leftrightarrow
 F_1(\pmb{\phi})=\sum_{j=1}^Ju_jh_j\exp(-H_j)\Delta=0=f_1 ~;~ \\
&
\V(\epsilon)=\sigma^2_\epsilon=1
 \Leftrightarrow
 F_2(\pmb{\phi})=\sum_{j=1}^Ju_j^2h_j\exp(-H_j)\Delta-F_1(\pmb{\phi})^2=1=f_2~.
\end{align*}
Let $\tilde{h}_j=\tilde{h}_\epsilon(u_j)$, $\tilde{H}_j=\sum_{\ell\leq j}\tilde{h}_j\Delta$,
$\tilde{f}_j=\tilde{h}_j\exp(-\tilde{H}_j)$ and $b_{jk}=b_k(u_j)$. Then, one can show that
%\begin{align*}
%&
$
\tilde{\mathbf{V}}_{1k}={\partial F_1(\tilde{\pmb{\phi}})\over \partial \phi_k}
=\sum_{j=1}^J u_j \tilde{f}_j \Delta \left(b_{jk}-\sum_{\ell\leq
    j}b_{\ell k}\tilde{h}_\ell \Delta\right)%,\\
$
%&
and
$
\tilde{\mathbf{V}}_{2k}={\partial F_2(\tilde{\pmb{\phi}})\over \partial \phi_k}
=\sum_{j=1}^J u_j^2 \tilde{f}_j \Delta \left(b_{jk}-\sum_{\ell\leq j}b_{\ell k}\tilde{h}_\ell \Delta\right)
 -2 F_1(\tilde{\pmb{\phi}})\,\tilde{\mathbf{V}}_{1k}.
$
%\end{align*}
Combining these last results with the elements from Sections
\ref{DensityEstRCData:Sec} and \ref{DensityEstICData:Sec}, one can
estimate the spline parameters underlying the hazard and, hence, the
density, for given (potentially) right- or interval-censored data and
penalty parameter $\tau$. The following section is devoted to the
selection of $\tau$.

\subsubsection{Selection of the penalty parameter $\tau$} \label{SelectionPenaltyParmTau:Sec}
Given the following priors,
\begin{align}
\tau\sim\G{1}{b}~~;~~p(\pmb{\phi}|\tau)\propto \tau^{K-r\over 2}
\exp\left(-{\tau\over 2}\,\pmb{\phi}^\top\mathbf{P}\pmb{\phi}\right),
\label{DensityPriors:Eq}
\end{align}
the joint posterior for the spline and the penalty parameters
$(\pmb{\phi},\tau)$ are
\begin{align}
p(\pmb{\phi},\tau|D) \propto
 \exp\{\ell(\pmb{\phi}|\D)\}\,p(\pmb{\phi}|\tau)\,p(\tau)
=\exp\{\ell_p(\pmb{\phi}|\tau,\D)\}\,\tau^{K-r\over 2}\,p(\tau).
\label{DensityJointPost:Eq}
\end{align}
Using the same arguments as in Section \ref{MarginalPostLambda:Sec} for
$(\pmb{\psi}|\pmb{\lambda},D)$,
the conditional posterior for the spline parameters,
$
p(\pmb{\phi}|\tau,D) \propto \exp\{\ell_p(\pmb{\phi}|\tau,\D)\},
$
can be shown to be approximately %Gaussian,
\begin{align}
(\pmb{\phi} | \tau,\D)
\mathrel{\dot\sim}
\N{\hat{\pmb{\phi}}_\tau}{\hat\Sigma_\tau},
\label{ApproxConditionalPhiGivenTau:Eq}
\end{align}
where $\hat{\pmb{\phi}}_\tau$ denotes the conditional posterior mode
(equal to the penalized MLE of $\pmb{\phi}$ given $\tau$, see Sections
\ref{DensityEstRCData:Sec} and \ref{DensityEstICData:Sec}),
$
\hat\Sigma_\tau^{-1}
= \mathbf{H}_{\tau}(\hat{\pmb{\phi}}_\tau)
= \mathbf{B}^\top {\mathbf{W}}_\tau \mathbf{B} + \tau\mathbf{P}
$, cf.\,Eq.\,(\ref{HesLpen:Eq}),
with
$\mathbf{W}_\tau=\diag(\mathbf{w}_\tau)$,
$\mathbf{w}_\tau=\mathbf{n}\hat{{\mathbf{h}}}_\tau\Delta$ and
$\hat{\mathbf{h}}_\tau$ giving the estimated hazard at the bin
midpoints when $\pmb{\phi}=\hat{\pmb{\phi}}_\tau$. Given that the number
of observations $(\mathbf{k})_j$ in bin ${\cal J}_j$ has expected
value $(\mathbf{w})_j=(\mathbf{nh}\Delta)_j$, one might reasonably
approximate the last variance-covariance matrix by
$
\hat\Sigma_\tau^{-1}
\approx
\mathbf{B}^\top \mathbf{W} \mathbf{B}
+ \tau\mathbf{P}
$
with $\mathbf{W}=\diag(\mathbf{k})$,
thereby restricting its explicit dependence on $\tau$ to the
$\tau\mathbf{P}$ term.
The marginal posterior for $\tau$ is given by
\begin{align}
p(\tau|\D) =
{p(\pmb{\phi},\tau|\D)
\over
p(\pmb{\phi}|\tau,\D)
}
\mathrel{\dot\propto}
p(\hat{\pmb{\phi}}_\tau,\tau|\D)\,
|\mathbf{B}^\top \mathbf{W} \mathbf{B} + \tau\mathbf{P}|^{-1/2}
\label{MarginalTau:Eq1}
\end{align}
with the approximation coming from
(\ref{ApproxConditionalPhiGivenTau:Eq}) and the substitution of
$\mathbf{W}_\tau$ by $\mathbf{W}$.  Now consider a singular value
decomposition of penalty matrix,
$\mathbf{P}= \mathbf{U} \pmb{\Upsilon} \mathbf{U}^\top$, where
$\mathbf{U}=[\mathbf{U}_1~ \mathbf{U}_0]$,
$\mathbf{U}^\top\mathbf{U}=\mathbf{I}_K$,
$\pmb{\Upsilon}=\mathrm{blockdiag}(\pmb{\Upsilon}_1, \mathbf{0}_r)$,
with the last $r$ diagonal elements of
$\pmb{\Upsilon}=\diag(\pmb{\upsilon})$ being zero.
Then, using properties of determinants and defining
$\tilde{\mathbf{B}}=\mathbf{W}^{1/2}\mathbf{B}\mathbf{U}$,
$\tilde{\mathbf{B}}_1=\mathbf{W}^{1/2}\mathbf{B}\mathbf{U}_1$,
$\tilde{\mathbf{B}}_0=\mathbf{W}^{1/2}\mathbf{B}\mathbf{U}_0$,
$\mathbf{M}=\tilde{\mathbf{B}}_1^\top \tilde{\mathbf{B}}_1
 -\tilde{\mathbf{B}}_1^\top \tilde{\mathbf{B}}_0
   (\tilde{\mathbf{B}}_0^\top \tilde{\mathbf{B}}_0)^{-1}
   \tilde{\mathbf{B}}_0^\top \tilde{\mathbf{B}}_1
$,
one has
\begin{align}
&|\mathbf{B}^\top \mathbf{W} \mathbf{B} + \tau\mathbf{P}|
= |\tilde{\mathbf{B}}_0^\top \tilde{\mathbf{B}}_0|\,
|\mathbf{\Upsilon}_1|\,
\tau^{K-r}
\prod_{j=1}^{K-r} \left(1+{n \tilde{m}_j\over \tau}\right)
\label{DeterminantInvSigmaGivenTau:Eq}
\end{align}
where
$\widetilde{\mathbf{M}}
={1\over n}\mathbf{\Upsilon}_1^{-1/2}\mathbf{M}\mathbf{\Upsilon}_1^{-1/2}$
has eigenvalues $\{\tilde{m}_j\}_{j=1}^{K-r}$ independent of $\tau$.
Combining (\ref{DensityPriors:Eq}), (\ref{DensityJointPost:Eq}), (\ref{MarginalTau:Eq1}) and
(\ref{DeterminantInvSigmaGivenTau:Eq}), one has
\begin{align}
&\log p(\tau|\D)
  \mathrel{\dot =}
\ell_p(\hat{\pmb{\phi}}_\tau|\tau,\D)
+ \log p(\tau) -{1\over 2} \sum_{j=1}^{K-r} \log\left(1+{n
   \tilde{m}_j\over \tau}\right) \nonumber\\
 &~~= \ell(\hat{\pmb{\phi}}_\tau|\D)
  -\tau \left(b+ {1\over 2}\,\hat{\pmb{\phi}}_\tau^\top\mathbf{P}\hat{\pmb{\phi}}_\tau\right)
  -{1\over 2} \sum_{j=1}^{K-r} \log\left(1+{n\tilde{m}_j\over \tau}\right),
\label{MarginalTauLogPost:Eq}
\end{align}
suggesting Algorithm \ref{DensityEstimation:Algo} to select $\tau$.

%% Algorithm
% \begin{figure}[ht]
% \begin{Frame}[{Algorithm}: {\it Density estimation (selection of $\tau$ and
%   computation of $\hat{\pmb{\phi}}_\tau$)}]
\begin{algo} \label{DensityEstimation:Algo}
{\it Density estimation (selection of $\tau$ and computation of $\hat{\pmb{\phi}}_\tau$)}\\
{\it Principle}\,: the algorithm alternates the following two steps till
convergence:
\vspace{-\topsep}
\begin{enumerate}\itemsep0em
\item For a given value of the penalty parameter
$\tau$, select the spline parameters $\pmb{\phi}$ to maximize
$p(\pmb{\phi}|\tau,D)$ under the moments constraints $\E(\epsilon)=0$ and
$\V(\epsilon)=1$ ;
\item Update $\tau$ to maximize the approximation
(\ref{MarginalTauLogPost:Eq}) to $\log p(\tau|D)$.
\end{enumerate}
{\it Practically}\,: repeat till convergence:
\vspace{-\topsep}
\begin{enumerate}\itemsep0em
\item Given the current estimate for $\tau$, maximize the Lagrangian in
  (\ref{DensityEstimation:Lagrangian:Eq}) by repeating the
  Newton-Raphson step in (\ref{ConstrainedDensityEst:Eq}) till
  convergence to $\hat{\pmb{\phi}}_\tau$.
\item Update $\tau$ by using the
fixed-point method on the partial derivative of
$(\ref{MarginalTauLogPost:Eq})$ w.r.t.\,$\tau$ set to zero. Practically, repeat
till convergence
$$
\tau ~\longleftarrow
{\sum_{j=1}^{K-r} {n\tilde{m}_j \over \tau+n\tilde{m}_j}
 /
 \left(2b+\hat{\pmb{\phi}}_\tau^\top\mathbf{P}\hat{\pmb{\phi}}_\tau\right).
}
$$
\end{enumerate}
At convergence, it yields $(\hat\tau,\hat{\pmb{\phi}}=\hat{\pmb{\phi}}_{\hat\tau})$
and the estimated hazard $\hat{h}_\epsilon(\cdot) =
\exp\Big(\sum_{k=1}^K b_k(\cdot) \hat{\phi}_k\Big)$.
\end{algo}
For example, with a dataset of size $n=1\,000$ including 40\%
uncensored, 40\% interval-censored and 20\% right-censored data, the
selection of $\tau$ and the estimation of $K=50$ B-spline parameters (an
unnecessary very large $K$ used to challenge Algorithm
\ref{DensityEstimation:Algo})
took 6 iterations and one tenth of a second using pure R code on a
small desktop computer.

\subsection{Algorithm for fitting the NP additive location-scale model}
\label{NPadditiveLocationScaleModel:Sec}
We now have all the necessary ingredients for fitting the
nonparametric double additive location-scale model (NP-DALSM) from
possibly right- or
even interval-censored data. The algorithm is
iterative and alternates the estimation of the error density (Step 1),
of the regression and spline parameters in the location (Step
2) and dispersion (Step 3)  submodels, selection of the penalty
parameters for the additive terms in location and dispersion (Step 4),
see Algorithm \ref{GlobalAlgorithm:Algo}.

%% Algorithm
% \begin{figure}
% {%\small
% \begin{Frame}[{Global Algorithm}: {\it Fitting the NP additive
%     location-scale model}]
\begin{algo} \label{GlobalAlgorithm:Algo}
\textrm{Global Algorithm}: {\it Fitting the NP additive
    location-scale model}\\
Iterate the following steps till convergence:
\vspace{-\topsep}
\begin{enumerate} \itemsep0em
\item {\it Estimation of the error hazard and density}:
\vspace{-\topsep}
 \begin{enumerate} \itemsep0em
   \item Given the current estimates for the regression and splines
     parameters, compute the standardized residuals
     $r_i={y_i-\mu_i(\pmb{\psi}^\mu_\lambda)\over\sigma_i(\pmb{\psi}^\sigma_\lambda)}$
     accompanied by their observation indicators $d_i$, some of these residuals
     being right-censored ($d_i=0$) or interval-censored with $r_i\in (r_i^L,r_i^R)$.
   \item Use Algorithm \ref{DensityEstimation:Algo} on these data
     to update the estimates of the error hazard
     function $h_\epsilon$ and density $f_\epsilon$. It is based on the
  estimation and selection of the underlying spline parameters
  $\pmb{\phi}$ and penalty parameter $\tau$.
 \end{enumerate}

\item {\it Estimation of $\pmb{\psi}^\mu$}: given the current values
  of the other parameters and in particular of the penalty parameter
  vector $\pmb{\lambda}^\mu$ for the additive terms in the
  location submodel, $\pmb{\psi}^\mu = \begin{pmatrix}\pmb{\beta},
    \Vec{\mathbf{\Theta}^\mu}\end{pmatrix}$ is reevaluated to maximize
  $
  p(\pmb{\psi}^\mu|\pmb{\psi}^\sigma,\pmb{\lambda}^\mu,\pmb{\phi},\D)
  \propto L(\pmb{\psi}^\mu,\pmb{\psi}^\sigma,\pmb{\phi};\D)
   ~p(\pmb{\psi}^\mu|\pmb{\lambda}^\mu)
  $
  using the Newton-Raphson (N-R) procedure described in
Algorithm \ref{PsiMuPsiSigma:Algo}
with the current estimate for
  $\pmb{\psi}^\mu$ as starting value.

\item {\it Estimation of $\pmb{\psi}^\sigma$}: given the current values
  of the other parameters and in particular of the penalty
  parameter vector $\pmb{\lambda}^\sigma$ for the additive terms in the
  dispersion submodel, $\pmb{\psi}^\sigma = \begin{pmatrix}\pmb{\delta},
    \Vec{\mathbf{\Theta}^\sigma}\end{pmatrix}$ is reevaluated to maximize
  $
  p(\pmb{\psi}^\sigma|\pmb{\lambda}^\mu,\pmb{\lambda}^\sigma,\pmb{\phi},\D)
  \propto L(\tilde{\pmb{\psi}}^\mu,\pmb{\psi}^\sigma,\pmb{\phi};\D)
  \,p(\pmb{\psi}^\sigma|\pmb{\lambda}^\sigma)
  \begin{vmatrix} \tilde{\Sigma}^\mu_\lambda \end{vmatrix}^{1/2}
  $
  using the Newton-Raphson (N-R) procedure described in
Algorithm \ref{PsiMuPsiSigma:Algo}
  with the current estimate for
  $\pmb{\psi}^\sigma$ as starting value.

\item {\it Selections of $\pmb{\lambda}^\mu$ and $\pmb{\lambda}^\sigma$}:
the penalty parameters in the additive terms are chosen to maximize
$\log\tilde{p}(\pmb{\lambda}^\mu|\pmb{\lambda}^\sigma,\D)$ and
$\log\tilde{p}(\pmb{\lambda}^\sigma|\pmb{\lambda}^\mu,\D)$
using Algorithm \ref{LambdaMuLambdaSigma:Algo}.
\end{enumerate}
Possible starting values are obtained by:
\vspace{-\topsep}
\begin{enumerate}[-]
\itemsep0em
\item Assuming a Gaussian error distribution ;
\item Discarding right-censored data and setting interval-censored
  ones to their midpoint value, yielding a reduced response vector
  $\tilde{\pmb{y}}$ with an associated design matrix
    $\tilde{\cal{X}}^\mu$ for the additive location submodel ;
\item Setting the elements in penalty vectors $\pmb{\lambda}^\mu$ and
  $\pmb{\lambda}^\sigma$ to a moderately large value (100, say)\,;
\item Estimating $\pmb{\psi}^\mu$ using penalized LS:
$\pmb{\psi}^\mu \longleftarrow
\left({{\tilde{\cal{X}}^{\mu^\top}}\tilde{\cal{X}}^\mu+\mathbf{K}^\mu_\lambda}\right)^{-1}
{\tilde{\cal{X}}^{\mu^\top}}\tilde{\pmb{y}}$~;
\item Fixing $\pmb{\psi}^\sigma$ to zero, except its first component
  $\delta_0$ set to the log of the mean squared error.
\end{enumerate}
\end{algo}
Convergence is very fast with the suggested initial conditions. One
major advantage of our proposal is that it does not require
backfitting as regression and spline parameters are updated
simultaneously within the location and dispersion submodels. An
additional remarkable feature is the joint update of the (log of the)
penalty parameters using a Newton-Raphson procedure based on
approximate analytical expressions for the gradient and Hessian of
their marginal posterior. And last but not least, the error
distribution is also estimated through the underlying (log-)hazard
expressed as a linear combination of (penalized) P-splines with a
penalty parameter selected to maximize its posterior density.  The
whole procedure is able to handle right- or interval-censored response
data.

%%%
\section{Simulation Study} \label{SimulationStudy:Sec}
An extended simulation study was made to evaluate the
performances of the proposed algorithm to fit the nonparametric
additive location-scale model. The data were simulated with
conditional location and dispersion given by, respectively,
\begin{align}
\mu(\mathbf{z}^\mu, \mathbf{x}^\mu)
&= (\beta_{0}+\beta_1z_1^\mu+\beta_2z_2^\mu)
  +f_{1}^{\mu}(x_1^\mu)+f_{2}^{\mu}(x_2^\mu), \label{SimulMean:Eq}\\
\log\sigma(\mathbf{z}^\sigma, \mathbf{x}^\sigma)
&= (\delta_{0}+\delta_1z_1^\sigma+\beta_2z_2^\sigma)
  +f_{1}^\sigma(x_1^\sigma)+f_{2}^\sigma(x_2^\sigma). \label{SimulDisp:Eq}
\end{align}
Different combinations of
sample sizes $n$ ($=1500, 500, 250$), right censoring (RC
$=0\%,25\%,50\%$) rates and interval censoring (IC $=0\%,25\%,50\%$) rates
were considered. The standardized error term (with mean 0 and variance
1) in (\ref{LocationScaleModel:Eq}) was taken to have a Normal
mixture distribution,
% \begin{align}
% &
$\epsilon \sim .8~\N{-0.414}{0.538^2} + .2~\N{1.655}{0.646^2}$,
% \label{SimulErrorDist:Eq}
% \end{align}
see Fig.\,\ref{FittedErrorDensities:Fig} in the Supplementary Material.
For each of the $n$ units, the pair of covariates
($p_1=p_2=2$) with linear effects in (\ref{SimulMean:Eq}) and
(\ref{SimulDisp:Eq}) were independently generated from Bernoulli and
Normal distributions,
$z_1^\mu, z_1^\sigma \sim \mathrm{Bern(.6)}~;~z_2^\mu,z_2^\sigma\sim\N{0}{1},$
with regression parameters
$\pmb{\beta}=(1.6,.3,.75)$, $\pmb{\delta}=(-.5,-.03,.01)$.
Two $(=J_1=J_2)$ additive terms per regression submodel were added,
% \begin{equation*}
% \begin{array}{ll}
$f_1^\mu(x) = .113-.4\sqrt{x}\sin(1.2\pi x)$,
%&;~
$f_2^\mu(x) = .586 -.3(x^2+.3)^{-1}$, % \\
$f_1^\sigma(x) = -0.158 + 0.15x + 0.25x^2$, % &;~
$f_2^\sigma(x) = 12 (x - 0.5)^3$,
% \end{array}
% \end{equation*}
with $x_1^\mu,x_2^\mu,x_1^\sigma,x_1^\sigma$ generated independently
from a uniform distribution on $(0,1)$, see the solid curves on
Fig.\,\ref{FittedAdditiveTermsN250-IC0:Fig} in the Supplementary
Material for a graphical representation.  For each of the $n$ units,
covariates were first sampled to define the underlying first and
second order (conditional) moments in (\ref{SimulMean:Eq}) and
(\ref{SimulDisp:Eq}), yielding $\mu_i$ and $\sigma_i$ for the $i$th
unit. The associated uncensored response was then obtained using
$y_i=\mu_i+\sigma_ie_i$ with $e_i$ sampled from the Normal
mixture. %(\ref{SimulErrorDist:Eq}).
Right censoring was created randomly and independently of the
underlying response and covariates using an exponential distribution
$C_i\sim\Exp(\lambda)$ with $\lambda$ selected to reach the desired
percentage $RC$ of right censored responses. The observed response was
then defined as $t_i=\min\{y_i,c_i\}$ with {\it observation} indicator
$\delta_i=I(c_i>y_i)$. The non right-censored data (for which
$\delta_i=1$) were subsequently interval-censored with probability
$IC/(1-RC)$ with, then, $y_i$ only reported to lie in $(y_i^L,y_i^R)$
where $y_i^L=y_i-1.5u_i\sigma(Y)$ and $y_i^R=y_i+1.5(1-u_i)\sigma(Y)$
with $u_i\sim U_{(0,1)}$, yielding an interval of width equal to 1.5
the marginal standard deviation of the response.

The double additive location-scale model (DALSM) was fitted by
assuming a nonparametric (NP) or a Normal ($\cal N$) density for the
error term. Under the working Normality hypothesis, the sandwich
estimator \citep{White1982} was preferred over the model-based one for
the variance-covariance of the regression and spline parameter
estimates. A report on the detailed simulation results can be found in
Supplementary Material \ref{Appendix:C}.  In summary, our simulation
study suggests that the proposed NP estimation strategy enables to
quantify the effects of covariates on location and dispersion with
negligible biases and important efficiency gains as compared to an
approach assuming normality.  Uncertainty in the estimation is
properly quantified, except when the sample size is small (as compared
to the number of parameters in the model). Then, the effective
coverage of credible intervals can be smaller than the nominal
value. In these cases, MCMC with proposals built using approximated
posteriors resulting from the algorithm in Section
\ref{NPadditiveLocationScaleModel:Sec} would generate more reliable
quantification of uncertainty, but at a higher computational cost.
The error density is properly estimated in the absence of right
censoring even with a rather small sample size and a large interval
censoring rate. But the combination of a small $n$ and a large right
censoring rate somehow decrease the quality of the expected
reconstruction as the available information on the error distribution
becomes sparse and incomplete. Then, the smallest component in the
Normal mixture tends to be flattened around its mode.

%%%
\section{Application} \label{Applications:Sec}

The proposed application involves interval- and right-censored
responses. The data of interest come from the European Social Survey
\citep{ESS:2016}. We focus on the money available per person in
Belgian households for respondents aged 25-55 when the main source of
income comes from wages or salaries ($n=756$). Each person reports the
total net monthly income of the household in one of 10 decile-based
intervals: 1:\,$<1.120$ $(n_1=8)$, 2:\,$[1.120,1.399]$ $(n_2=13)$,
3:\,$[1.400,1.719]$ $(n_3=47)$, 4:\,$[1.720,2.099]$ $(n_4=53)$,
5:\,$[2.100,2.519]$ $(n_5=82)$, 6:\,$[2.520,3.059]$ $(n_6=121)$,
7:\,$[3.060,$ $3.739]$ $(n_7=167)$, 8:\,$[3.740,4.529]$ $(n_8=126)$,
9:\,$[4.530,5.579]$ $(n_9=74)$, 10:\,$\geq 5.580$ euros $(n_{10}=65)$.

We model the relation of the available income per person ($91.4\%$ are
interval-censored, $8.6\%$ right-censored) to the availability of
(at least) 2 salaries ($64.2\%$) in the household, the age
(\ttf{Age}: $41.0\pm8.83$ years) and the number of years of full-time education
completed (\ttf{Educ}: $14.9\pm3.34$ years) by the respondent.
That individualized income is obtained by dividing the household one by the
OECD-modified equivalence
scale \citep{Hagenaars1994}, as recommended by the Statistical Office of
the European Union (EUROSTAT). The first adult in the household
contributes to 1.0 to that scale, each person aged at least 14 adds .5
to it, while each younger member brings an extra .3 to the household
weight. For example, a respondent aged 31 declaring a household net
monthly income in the interval $(3060,3740)$ euros with a partner aged 34
and 4 children aged 15, 10, 9 and 3 would be associated to an OECD-modified
scale of 2.9 and an interval-censored response of $(1055.2,1289.7)$
euros (available per person).

The nonparametric double additive location-scale model (NP-DALSM)
described in Section \ref{AdditiveLocationScale:Sec} with the flexible
error density from Section \ref{NPDensityEstimation:Sec} was fitted
using Algorithm \ref{GlobalAlgorithm:Algo}: 10 (=$L$) and 20 (=$K$)
B-splines were taken to model the additive terms and the log hazard of
the error distribution, respectively.
The response was rescaled in
thousand euros, while quantitative covariates were relocated and
rescaled to take values in $(0,1)$ before running the algorithm. It
converged after 10 iterations in about 2 seconds using pure R
code.
\begin{figure}\centering
\begin{tabular}{cc}
\includegraphics[width=10cm]{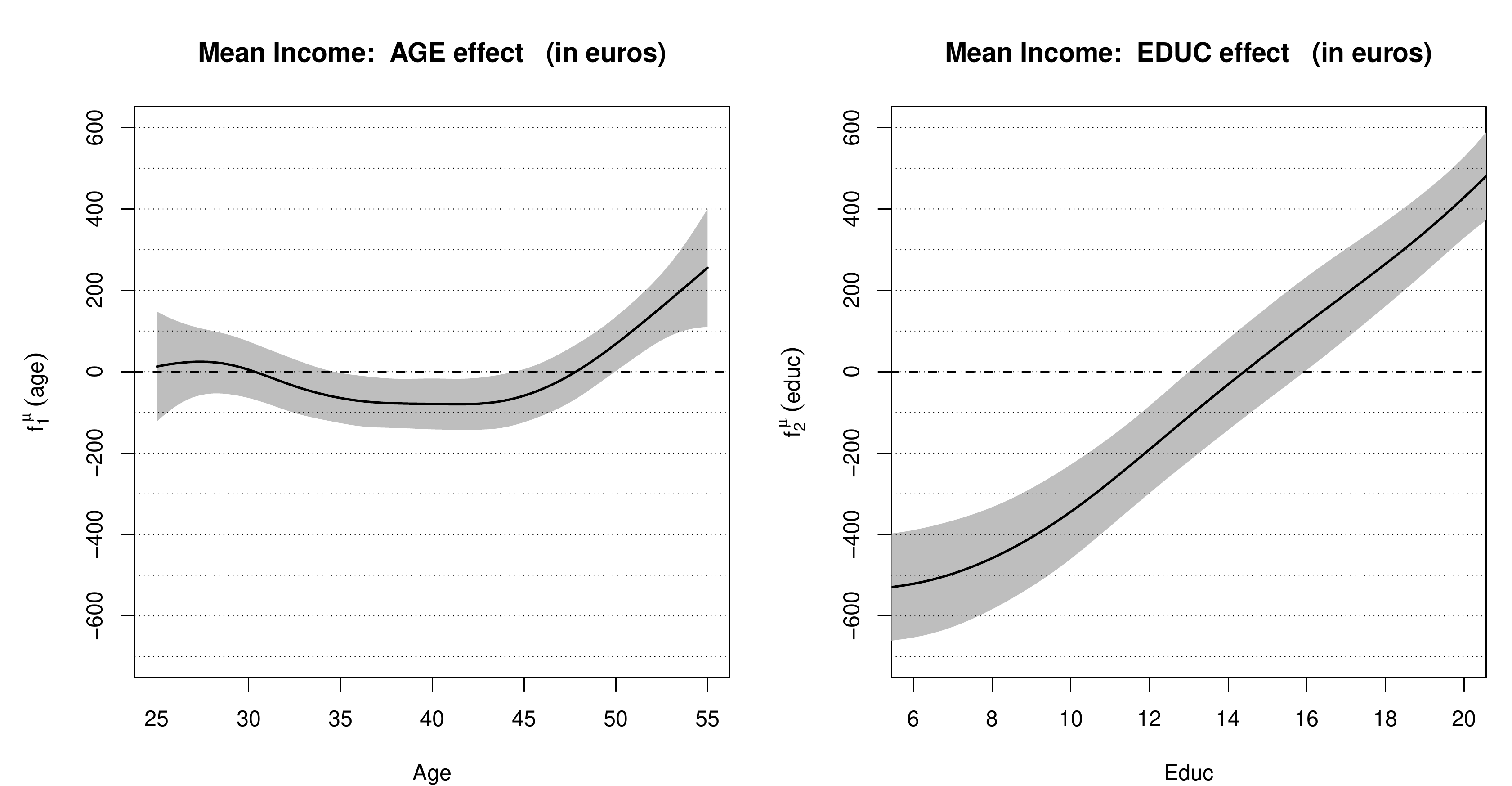}&\\
\includegraphics[width=10cm]{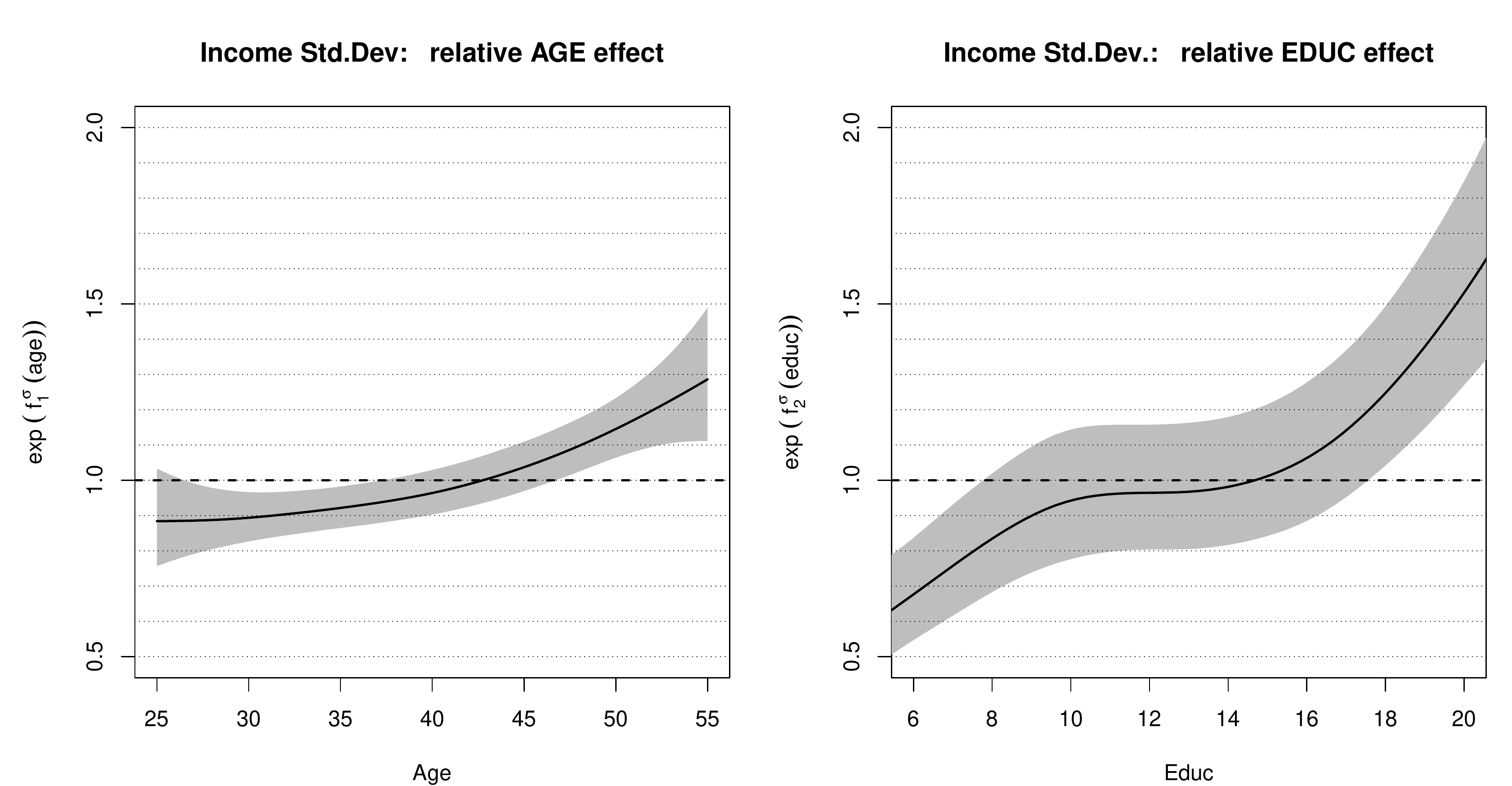}&\\
\includegraphics[width=6cm]{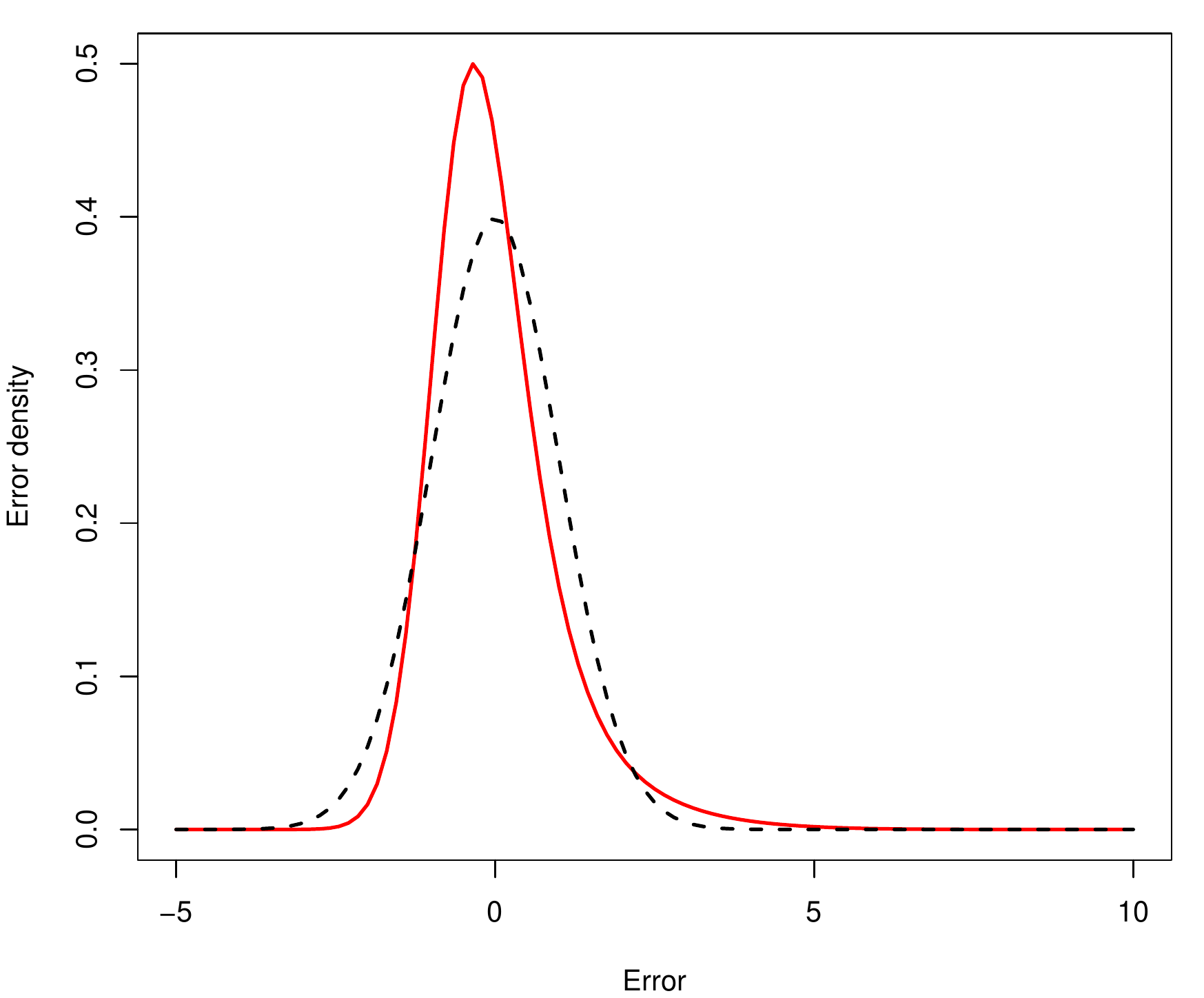}
\end{tabular}
\caption{Belgian income data (ESS 2016): estimated additive terms in the NP
  additive location-scale model with pointwise 95\% credible intervals
  ; Row\,1 (effects on location): $f_1^\mu(\text{Age})$ and $f_2^\mu(\text{Educ})$  in euros ;
  Row\,2 (relative effects on dispersion):
  $\exp\big(f_1^\sigma(\text{Age})\big)$ and
  $\exp\big(f_2^\sigma(\text{Educ})\big)$ ;
  Estimated error density (solid line) compared to the
  standard Normal (dashed line).
} \label{ESS2016:Fig}
\end{figure}
\begin{figure}\centering
\begin{tabular}{cc}
\includegraphics[width=6.5cm]{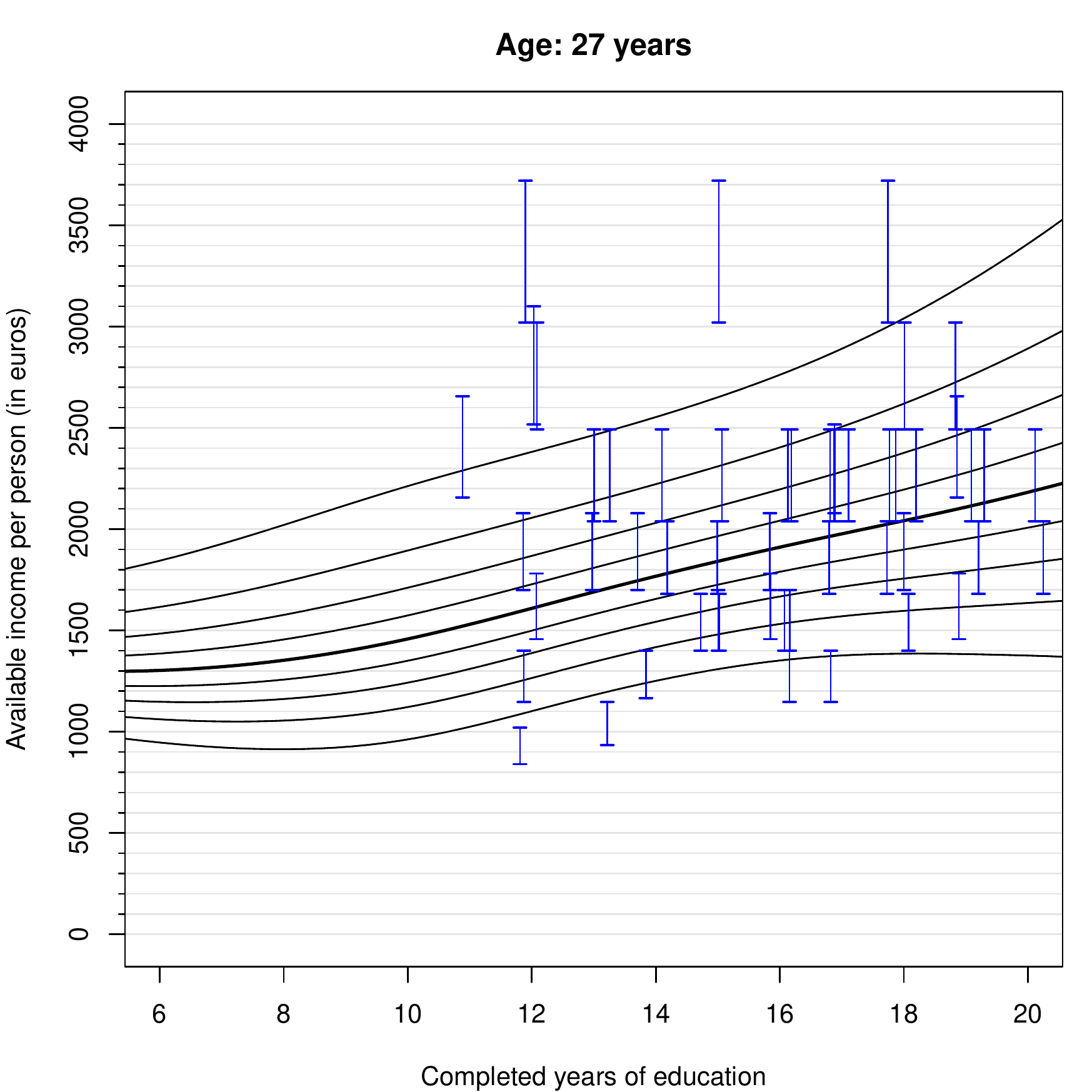}&
\includegraphics[width=6.5cm]{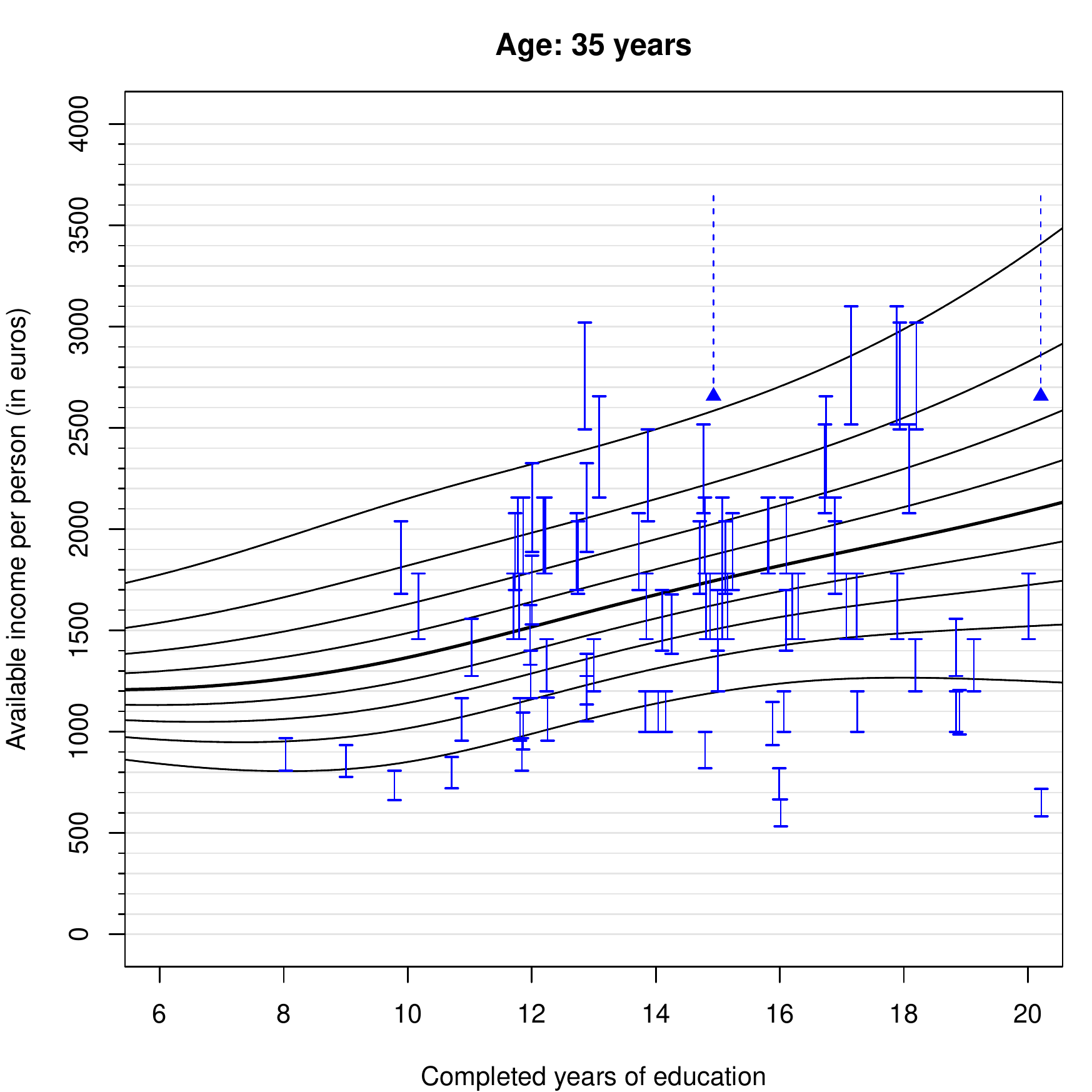}\\
\includegraphics[width=6.5cm]{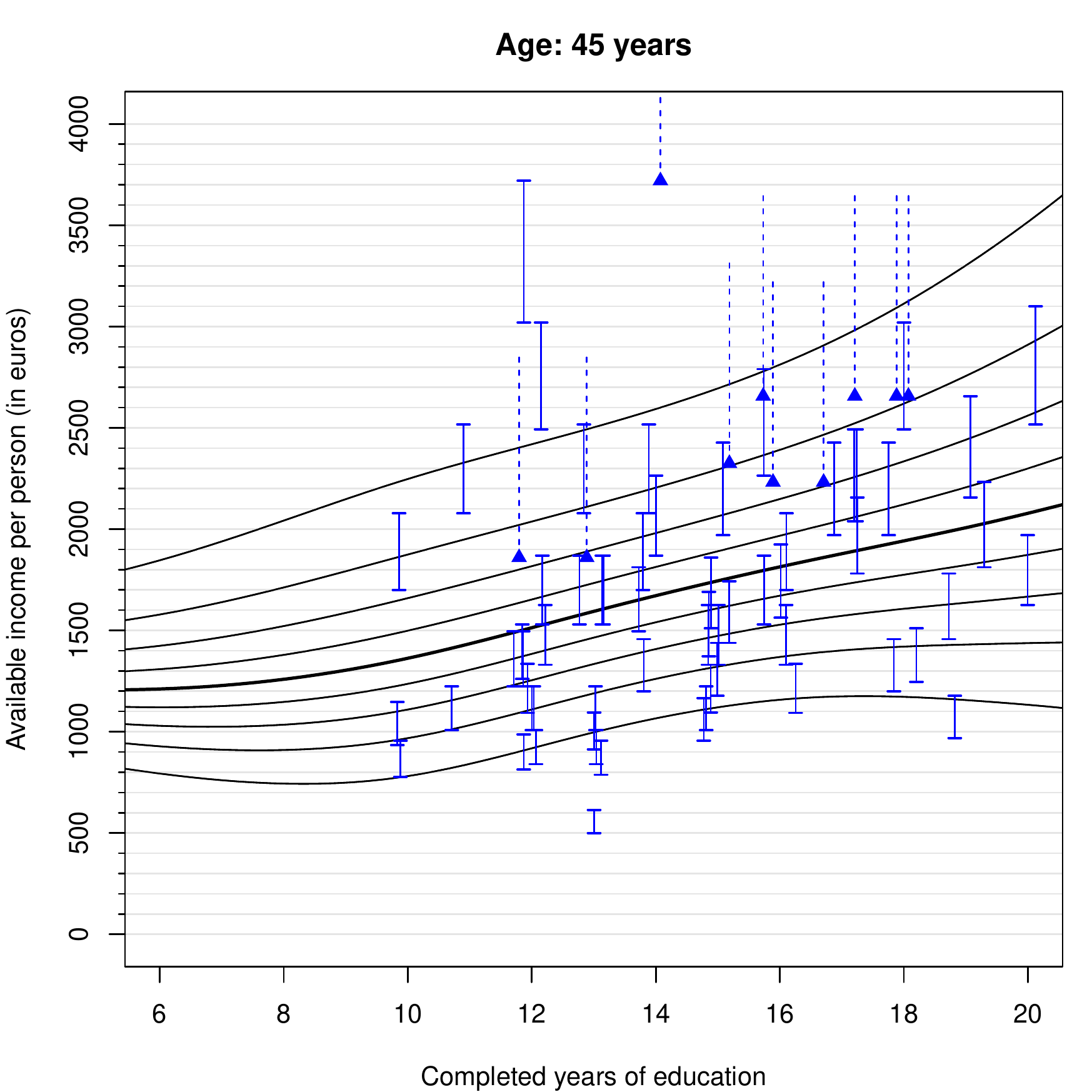}&
\includegraphics[width=6.5cm]{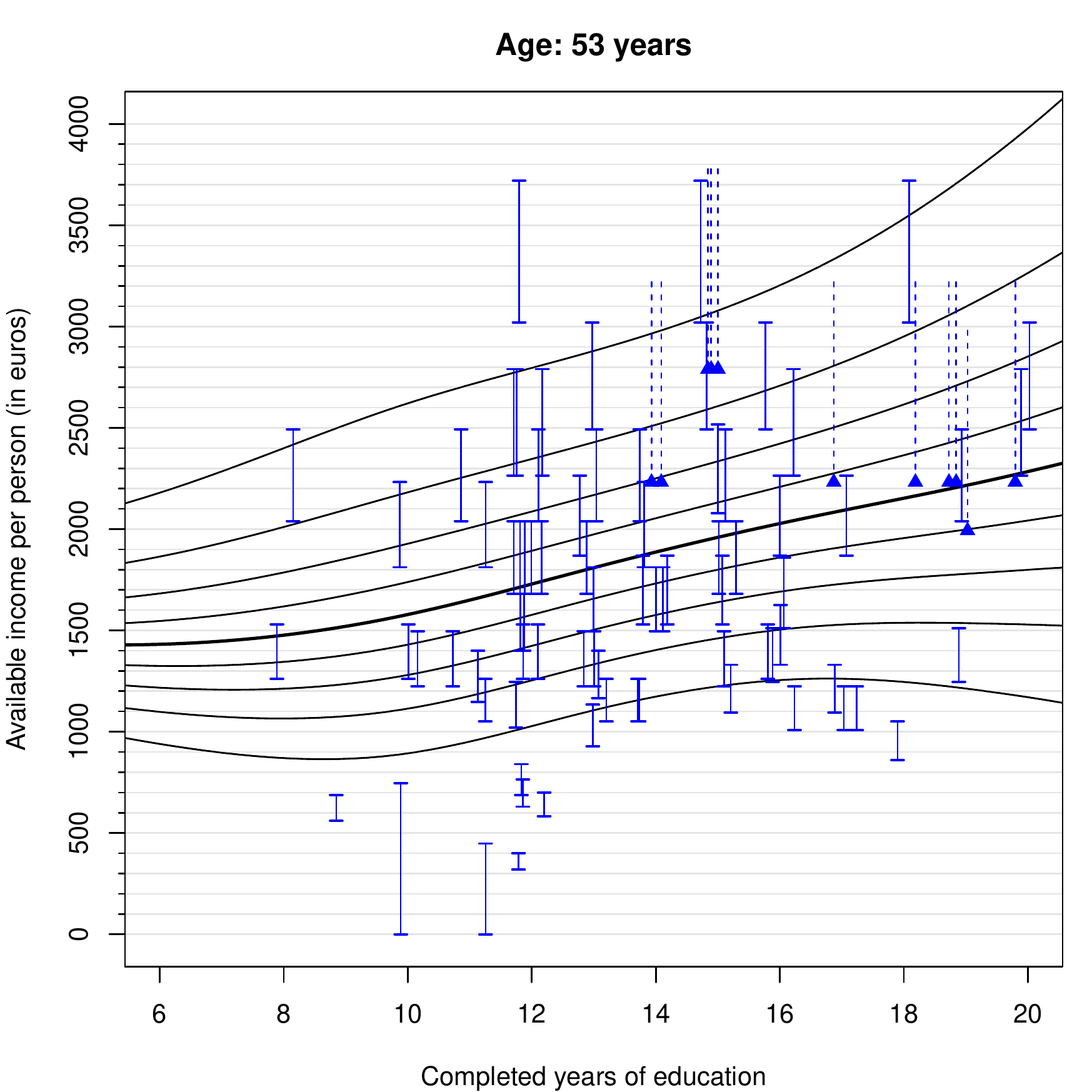}
\end{tabular}
\caption{Belgian income data (ESS 2016): fitted conditional deciles for the income
  per person in two-income households.
} \label{ESS2016:FittedDeciles:Fig}
\end{figure}
\begin{table}[h]

\begin{center}
{\small
\begin{tabular}{lccccccc}
%\multicolumn{3}{l}{{\it Fixed effects for Location}\,:}\\
Fixed      & \multicolumn{3}{c}{Location} && \multicolumn{3}{c}{Dispersion}\\
\cline{2-4}\cline{6-8}
effects &  $\hat{\beta}$ & s.e.   & CI 95\%       && $\hat{\delta}$ & s.e. & CI 95\% \\
\hline
\ttf{Intercept}  & 1.589 &0.057   &(1.478,~1.699) && -0.430 &0.091 & (-0.608, -0.251)\\
\ttf{TwoIncomes} & 0.266 &0.042   &(0.183,~0.349) && -0.020 &0.060 & (-0.137,~~0.097)\\
\end{tabular}
\vspace{.5cm}

%{\it Additive terms for Location}:\\
\begin{tabular}{lccccc}
%\multicolumn{3}{l}{{\it Additive terms for Location}\,:}\\
Additive   & \multicolumn{2}{c}{Location} && \multicolumn{2}{c}{Dispersion}\\
\cline{2-3}\cline{5-6}
terms      & e.d.f.&  CI 95\% && e.d.f.&  CI 95\% \\
\hline
\ttf{Age}  & 3.69 & (2.50,~5.16) && 2.40 & (1.26,~4.35) \\
\ttf{Educ} & 3.55 & (2.31,~4.97) && 3.86 & (2.55,~4.98) \\
\end{tabular}
}
\end{center}
\caption{Belgian income data (ESS 2016): fixed effect estimates and
  effective degrees of freedom (e.d.f.) (with 95\% credible intervals)
  for the additive terms
  in the NP double additive location-scale model.}
\label{ESS2016:FixedEffects:Tab}
\end{table}
Parameter estimates quantifying the effect of the \ttf{TwoIncomes}
binary indicator on the conditional mean and the log of the standard
deviation can be found in Table \ref{ESS2016:FixedEffects:Tab},
suggesting an average increase of 266 euros per person in the
household when two members of the household work (conditionally on
\ttf{Age} and \ttf{Educ}), while the effect on dispersion is not
statistically significant.  The effects of \ttf{Age} and \ttf{Educ} on
the conditional mean and dispersion can be visualized on the first and
second rows of Fig.\,\ref{ESS2016:Fig}, respectively, with the
corresponding estimated additive terms.  The money available per
household member tends to decrease with age (see
$f_1^\mu(\text{Age})$) between 25 and 40 (most likely due the arrival
of children in the family) and to increase afterwards (probably thanks
to wage increase with seniority and the departure of children). The
dispersion, reported as the exponential of the additive term,
$\exp(f_1^\sigma(\text{Age}))$, significantly increases with \ttf{Age}
with an acceleration over 45. However, the dominating effect comes
from the education level of the respondent with approximately a
difference of 1\,000 euros (in expected available income per person)
between a less educated (6 years) and a highly educated (20 years)
one, see $f_2^\mu(\text{Educ})$. The effect on dispersion is also
large, see $\exp(f_2^\sigma(\text{Educ}))$, with essentially an
important contrast between less and highly educated respondents, the
latter group showing the largest heterogeneity. Indeed, while most low
skilled persons have difficulties to find a job or are confined to
low-pay professions, a university degree offers a large variety of
opportunities from a moderately paid civil servant job to a manager
position in a multinational corporation in the chemical,
pharmaceutical or financial sectors.  The estimated density for the
error term can also be seen at the bottom of Fig.\,\ref{ESS2016:Fig},
with a right-skewed shape clearly distinguishable from the Gaussian
one typically assumed in parametric location-scale regression models.
The resulting estimates for the deciles of the income available per
person for varying education levels and ages are pictured on
Fig.\,\ref{ESS2016:FittedDeciles:Fig}. Interval- and right-censored
data are represented as intervals and dashed semi-intervals,
respectively (with horizontal noise added to untie respondents sharing
the same age). The precedingly discussed combined nonlinear impacts of
age and education level on the distribution of the available income
per person are now clearly visible.

%%%%
\section{Discussion} \label{Discussion:Sec}

The proposed nonparametric double-additive location-scale model
(NP-DALSM) is a fast and efficient alternative to parametric
location-scale models. Unlike moment-based estimation approaches such
as the generalized method of moments \citep[see e.g.][]{Wang2014}, it
provides a full estimation of the conditional distribution of the
response, that can be used to understand and visualize how it is
qualitatively and quantitatively affected by covariates. The density
of the error distribution is estimated from possibly right- or
interval-censored responses under moment constraints.  The penalty
parameters controlling the smoothness of the additive terms in the
location and dispersion submodels are automatically selected using
approximations to their marginal posteriors. These are obtained by
substituting Laplace approximations to the conditional posteriors of
the spline parameters, see Section \ref{MarginalPostLambda:Sec}.

Simulations suggest that the effects of covariates are properly
estimated with no significant biases in the estimation of regression
parameters and additive terms. The determinant in
(\ref{PsiSigmaMarginalApprox2:Eq}) plays an important role in the
process as its neglect would lead to non negligible biases in the
estimation of the dispersion part. Its role is comparable to the
correction brought by restricted maximum likelihood (REML) in more
elementary settings or in (adjusted) estimating functions, see e.g.\
\citet{JorgensenKnudsen2004}.  Biases in the estimation of the
intercepts can appear under large right censoring rates, while the
additive terms tend to be over-smoothed (as it should) when
information becomes sparse. It can for example result from the
combination of large right censoring rates and small sample sizes (as
compared to the large number of parameters to be estimated).

The nonparametric specification with P-splines of (the log-hazard
function underlying) the error density markedly increases the
efficiency of regression parameter and additive term estimates over
results under a working Normality hypothesis, and reduces the risk of
misleading conclusions following from a misspecified nonnormal
parametric density. While our proposal extends to nonparametric errors
and interval-censored settings some aspects of the remarkable work by
\citet{Wood:Fasiolo:2017} or \citet{wood2017generalized}, several
issues still need to be studied in that specific framework.  Model
validation is one topic, with the presence of interval-censored data
complicating the capacity to diagnose misspecification from partially
observed residuals. Model selection should also be
investigated. Obvious starting solutions would consist in computing
information criteria such as AIC and BIC with the number of parameters
replaced by effective dimensions \citep{Komarek2005}. The uncertainty
in the selection of the penalty parameters can also be accounted for,
see \citet{Wood2016b} or \citet[][Section 6.11]{wood2017generalized}
for additional perspectives. More elaborate procedures for testing the
necessity to include an additive term (in location or dispersion) or
to opt for a simpler linear form could be developed in our framework.
From a Bayesian perspective, they should be built using a combination
of the conditional posterior for the spline parameters of the additive
term of interest and the marginal posterior for the associated penalty
parameter.  Nonlinear and smooth interactions between covariates could
also be added to the location and dispersion parts in the same way as
\citet{Lee2011} and \citet{Rodriguez-Alvarez2018} with the conditional
mean in mixed models.

\appendix

\numberwithin{equation}{section}
\makeatletter
% "activate" the preparatory code, but for section-level headers only
\newcommand{\section@cntformat}{Appendix \thesection:\ }
\makeatother

% Appendix A
\section{Expressions for
$\pmb{\omega}^\mu, \pmb{\omega}^\sigma, \mathbf{W}^\mu,
\mathbf{W}^\sigma$} \label{Appendix:A}

Rewriting the error density as
$f_\epsilon(\cdot)=h_\epsilon(\cdot)\exp[-H_\epsilon(\cdot)]$ where
$H_\epsilon(\cdot)=-\log S_\epsilon(\cdot)$ and
$h_\epsilon(\cdot)=f_\epsilon(\cdot)/S_\epsilon(\cdot)$,
we obtain the following expressions (depending on the censoring status
of the response) for the elements of
$\pmb{\omega}^\mu$, $\pmb{\omega}^\sigma$  in $\rit^n$ and for
the diagonal elements
$\mathbf{w}^\mu$, $\mathbf{w}^\sigma$
in the $n\times n$ matrices $\mathbf{W}^\mu=\diag(\mathbf{w}^\mu)$,
$\mathbf{W}^\sigma=\diag(\mathbf{w}^\sigma)$:
\begin{description}
\item[Uncensored or right-censored $t_i$]: if $d_i$ is the censoring
  indicator, then
\begin{align} \label{OmegaW:UncensRightCens:eq}
\begin{split} % Make sure that a single number for the all set of equations
&\pmb{\omega}^\mu_i = -{1 \over \sigma_i}\left(d_i {h_i'\over h_i} -
  h_i\right) %, \\
%&
~;~\pmb{\omega}^\sigma_i = -d_ir_i {h_i'\over h_i} - d_i + r_ih_i~, \\
&\mathbf{w}^\mu_i = {1\over\sigma^2_i}
\left\{d_i\left(h_i'\over  h_i\right)^2
      -d_i{h_i''\over h_i}+h_i'\right\} , \\
&
\mathbf{w}^\sigma_i =
  d_i\left\{\left({h_i'\over h_i}\right)^2r_i^2 + {h_i'\over h_i}r_i - {h_i''\over h_i}r_i^2
     \right\}
  + h_i'r_i^2 + h_ir_i~,
\end{split}
\end{align}
where $h_i=h_\epsilon(r_i)$, $h_i'={dh_\epsilon(r_i)\over dr}$,
$h_i''={d^2h_\epsilon(r_i)\over dr^2}$ ;

\item[Interval-censored with $y_i\in(y_{i}^L,y_{i}^R)$]:
\begin{align}  \label{OmegaW:IntCens:eq}
\begin{split} % Make sure that a single number for the all set of equations
&\pmb{\omega}^\mu_i =
{1 \over \sigma_i} {f_\epsilon(r_i^L) - f_\epsilon(r_i^R) \over
  S_\epsilon(r_i^L)-S_\epsilon(r_i^R)} %\\
%&
~;~\pmb{\omega}^\sigma_i =
{r_i^Lf_\epsilon(r_i^L) - r_i^Rf_\epsilon(r_i^R) \over S_\epsilon(r_i^L)-S_\epsilon(r_i^R)}~,\\
&\mathbf{w}^\mu_i = {1\over\sigma^2_i}
\left[{f_\epsilon(r_i^L) g(r_i^L)- f_\epsilon(r_i^R) g(r_i^R) \over S_\epsilon(r_i^L)-S_\epsilon(r_i^R) }
+\left\{{f_\epsilon(r_i^L)-f_\epsilon(r_i^R) \over S_\epsilon(r_i^L)-S_\epsilon(r_i^R)}\right\}^2
\right], \\
&\mathbf{w}^\sigma_i =
\left\{{r_i^Lf_\epsilon(r_i^L) m(r_i^L)- r_i^Rf_\epsilon(r_i^R) m(r_i^R) \over S_\epsilon(r_i^L)-S_\epsilon(r_i^R) } \right\}
+\left\{{r_i^Lf_\epsilon(r_i^L)-r_i^Rf_\epsilon(r_i^R) \over S_\epsilon(r_i^L)-S_\epsilon(r_i^R)}\right\}^2 ,
\end{split}
\end{align}
where $g(r)={h_\epsilon'(r)/ h_\epsilon(r)}-h_\epsilon(r)$ and $m(r)=1+r g(r)$.
\end{description}

% Appendix B
\section{Gradient and Hessian of $E^\mu_\lambda$} \label{Appendix:B}
Denote the $i$th row of $\mathbfcal{X}^\mu$ (resp.\,$\mathbfcal{X}^\sigma$)
by the column vector $\mathbf{x}^\mu_i$ (resp.\,$\mathbf{x}^\sigma_i$).
Let us drop the  ``$\sim$'' sign to simplify notation and set
${\Sigma}^\mu_\lambda=\left({{\mathbfcal X}^\mu}^\top {\mathbf{W}}^\mu{\mathbfcal X}^\mu+\mathbf{K}^\mu_\lambda\right)^{-1}$.
One has
\begin{align*}
  & E^\mu_\lambda={1\over 2}\log\begin{vmatrix} {\Sigma}^\mu_\lambda \end{vmatrix}
  =-{1\over 2}
  \log \begin{vmatrix}
    \sum_{i=1}^n \mathbf{w}_i^\mu \mathbf{x}^\mu_i {\mathbf{x}^\mu_i}^\top+\mathbf{K}^\mu_\lambda
  \end{vmatrix}.
\end{align*}
Let
  $\mathbf{A}_k =
       {{\mathbfcal X}^\mu}^\top
         \diag\left(
                   {\partial \mathbf{w}^\mu \over \partial {\psi}_k^\sigma}
              \right)
   {\mathbfcal X}^\mu
   $
   and
   $
   \mathbf{A}_{k\ell} =
       {{\mathbfcal X}^\mu}^\top
         \diag\left(
                   {\partial^2 \mathbf{w}^\mu \over \partial {\psi}_k^\sigma\partial {\psi}_\ell^\sigma}
              \right)
       {\mathbfcal X}^\mu
   $
for $1\leq k,\ell \leq q_2$.
Reminding that for an arbitrary positive definite matrix $\mathbf{M}_t$,
 $
  {\partial \over \partial t}
  \log\begin{vmatrix}\mathbf{M}_t \end{vmatrix}
  =\mathrm{tr}\left(\mathbf{M}_t^{-1}
  {\partial \mathbf{M}_t\over \partial t}\right),
 $
 $
  {\partial \over \partial t}\mathbf{M}_t^{-1}
  = -\mathbf{M}_t^{-1} {\partial \mathbf{M}_t\over \partial t} \mathbf{M}_t^{-1},
 $
and using
 ${\partial \mathbf{w}_i^\mu / \partial {\psi}_k^\sigma}
  \approx -2\mathbf{w}_i^\mu\mathbf{x}_{ik}^\sigma,$
one can show that %one gets
%% Gradient
%%%%%%%%%%%
\begin{align*}
    {\partial E^\mu_\lambda\over \partial {\psi}_k^\sigma}
    &= -{1\over 2}
    \sum_{i=1}^n {\mathbf{x}^\mu_i}^\top
    {\Sigma}^\mu_\lambda
    \mathbf{x}^\mu_i \,{\partial \mathbf{w}_i^\mu \over \partial {\psi}_k^\sigma}
    = -{1\over 2}
    \mathrm{tr}
     \left(
        {\Sigma}^\mu_\lambda
        \mathbf{A}_{k}
     \right)
   \approx
    \sum_{i=1}^n \mathbf{w}_i^\mu
    \left({\mathbf{x}^\mu_i}^\top
         {\Sigma}^\mu_\lambda \mathbf{x}^\mu_i
    \right)
    \mathbf{x}^\sigma_{ik}\\
% %% Hessian
% %%%%%%%%%%
%   \begin{align*}
   -{\partial^2 E^\mu_\lambda\over \partial {\psi}_k^\sigma\partial{{\psi}_\ell^\sigma}}
   &= {1\over 2}
   \sum_{i=1}^n
   \left(
       {\mathbf{x}^\mu_i}^\top
    {\Sigma}^\mu_\lambda
    \mathbf{x}^\mu_i
   \right)
   \,{\partial^2 \mathbf{w}_i^\mu \over \partial {\psi}_k^\sigma \partial {{\psi}_\ell^\sigma}}
   -{1\over 2}
   \mathrm{tr}
   \left(
       {\Sigma}^\mu_\lambda
       \mathbf{A}_k
       {\Sigma}^\mu_\lambda
       \mathbf{A}_\ell
       \right)\\
  &
  = {1\over 2}
   \mathrm{tr}
   \left(
       {\Sigma}^\mu_\lambda
       \mathbf{A}_{k\ell}
   \right)
   -{1\over 2}
   \mathrm{tr}
   \left(
       {\Sigma}^\mu_\lambda
       \mathbf{A}_k
       {\Sigma}^\mu_\lambda
       \mathbf{A}_\ell
       \right)
\end{align*}

\section{Detailed simulation results} \label{Appendix:C}

\begin{figure}\centering
  {\small Location parameters ($n=1500$)}\\
  \includegraphics[width=12.5cm]{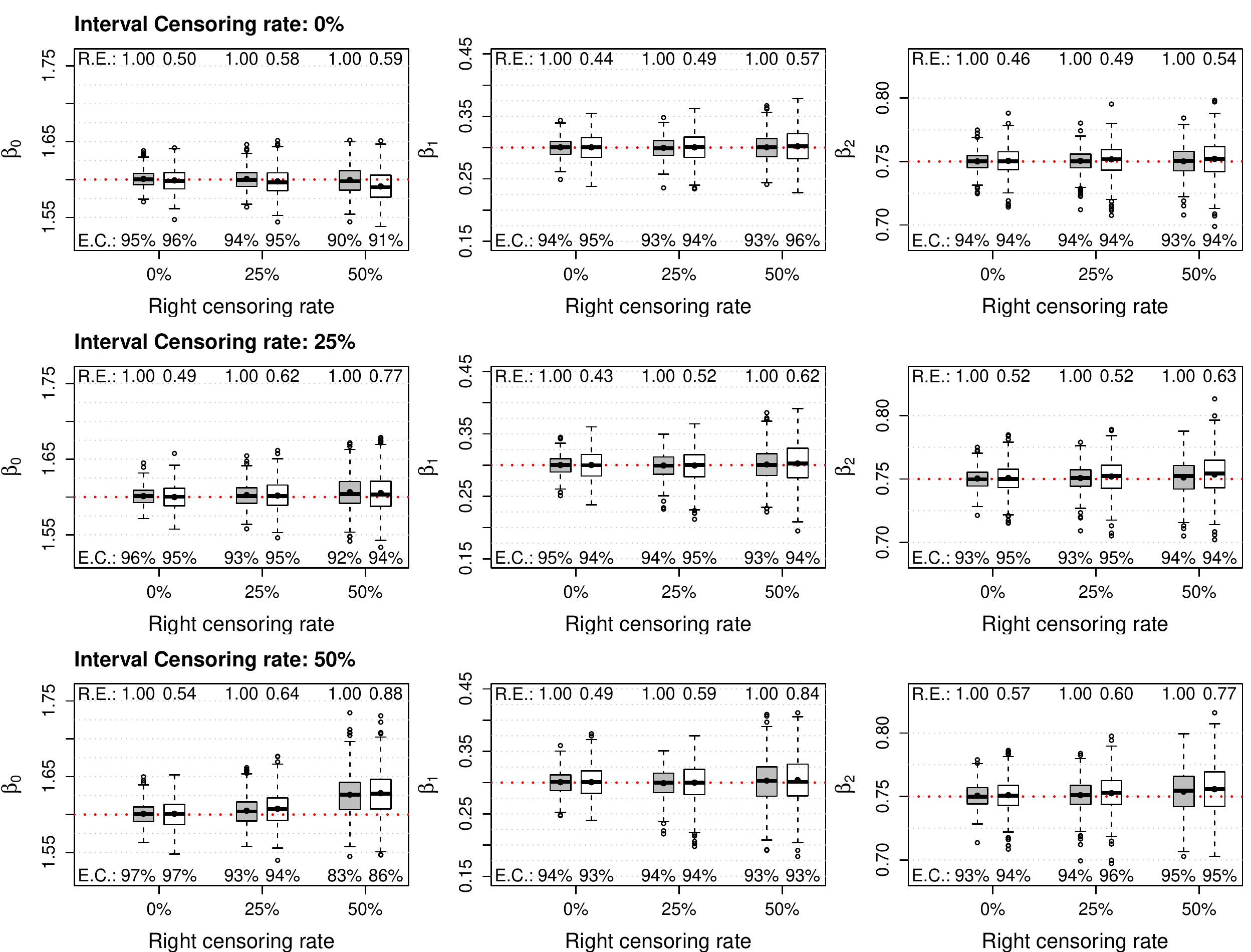}\\
  {\small Dispersion parameters ($n=1500$)}\\
  \includegraphics[width=12.5cm]{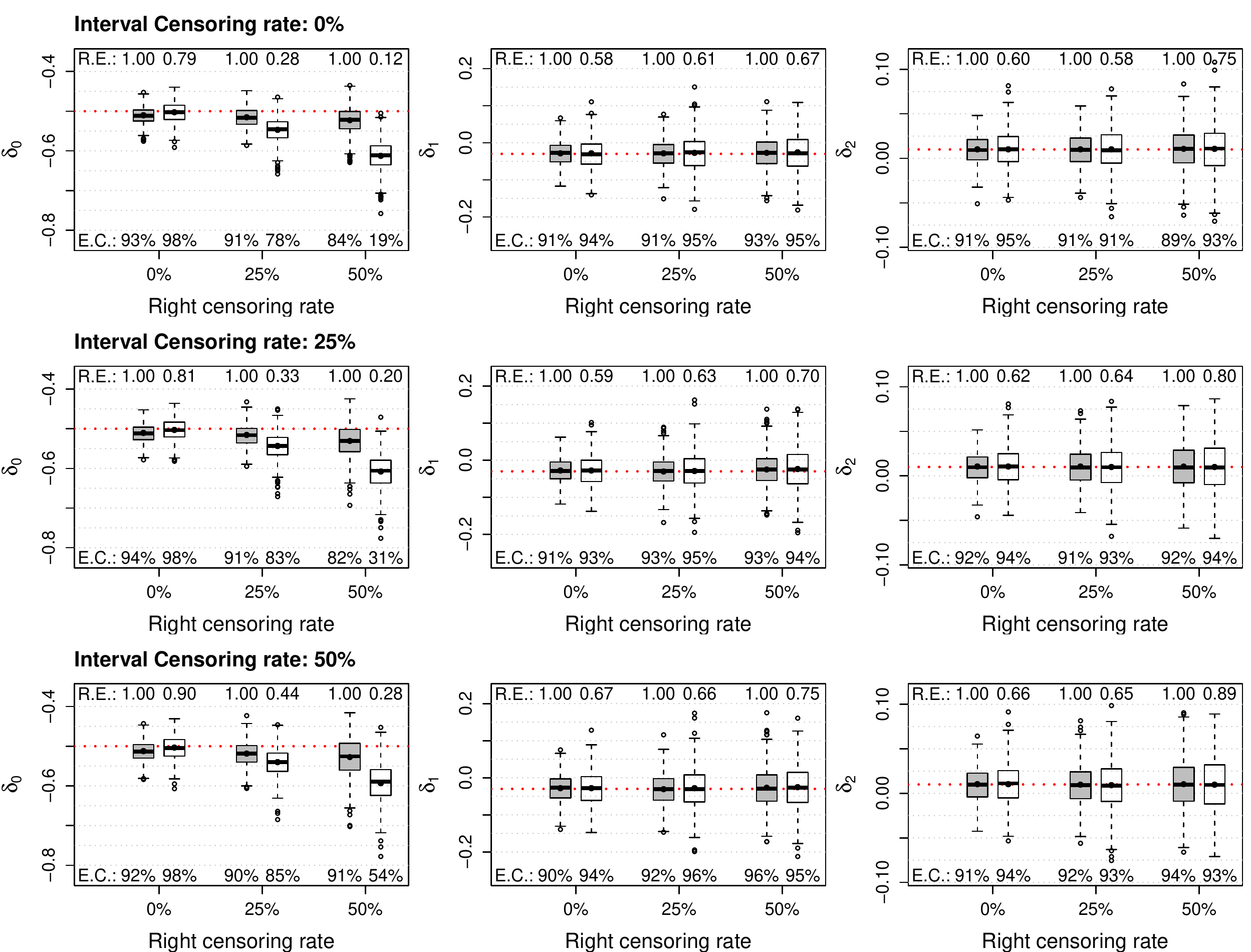}
\caption{Simulation study $(n=1500)$: estimation of the regression parameters in
  the double additive location-scale model over $S=500$ replicates:
  boxplot of the point estimates under a nonparametric (grey) or Normal (white)
  error term, Relative Efficiency (R.E.) under the working Normality
  hypothesis, Effective Coverage (E.C.) of 95\% credible intervals.
} \label{SimulationRegrParameters-n1500:Fig}
\end{figure}

\begin{figure}\centering
  {\small Location parameters ($n=500$)}\\
  \includegraphics[width=12.5cm]{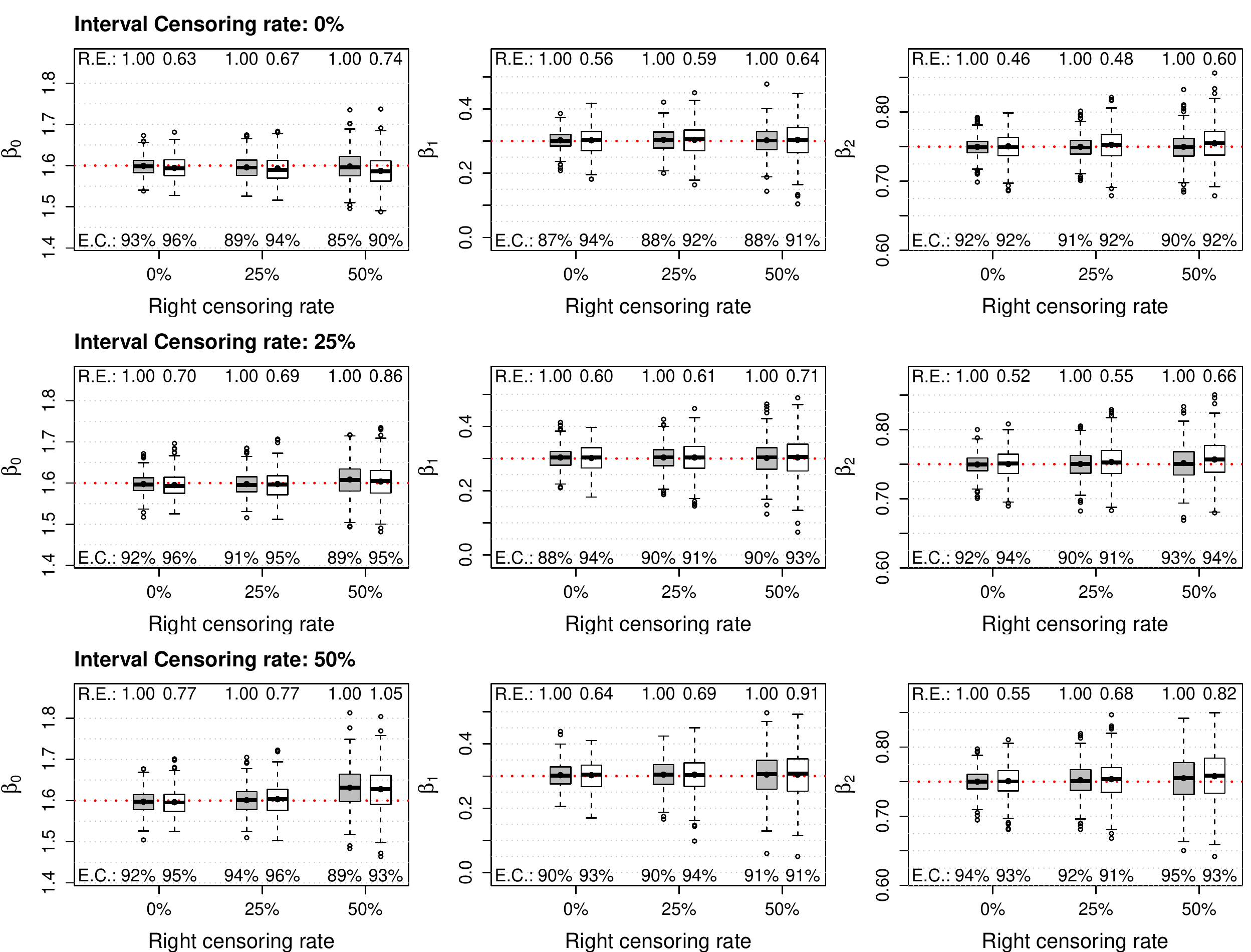}\\
  {\small Dispersion parameters ($n=500$)}\\
  \includegraphics[width=12.5cm]{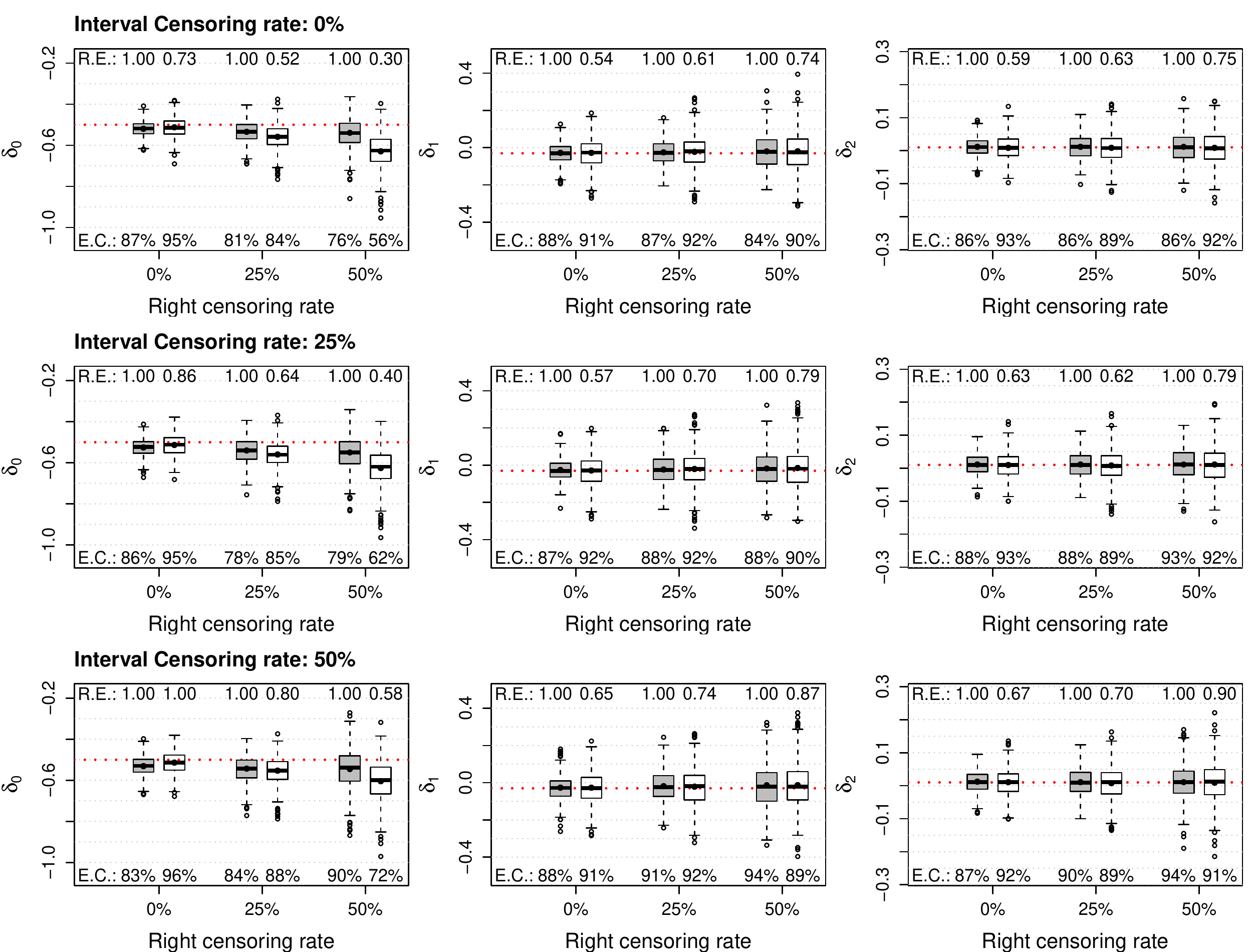}
\caption{Simulation study $(n=500)$: estimation of the regression parameters in
  the double additive location-scale model over $S=500$ replicates:
  boxplot of the point estimates under a nonparametric (grey) or Normal (white)
  error term, Relative Efficiency (R.E.) under the working Normality
  hypothesis, Effective Coverage (E.C.) of 95\% credible intervals.
} \label{SimulationRegrParameters-n500:Fig}
\end{figure}

\begin{figure}\centering
  {\small Location parameters ($n=250$)}\\
  \includegraphics[width=12.5cm]{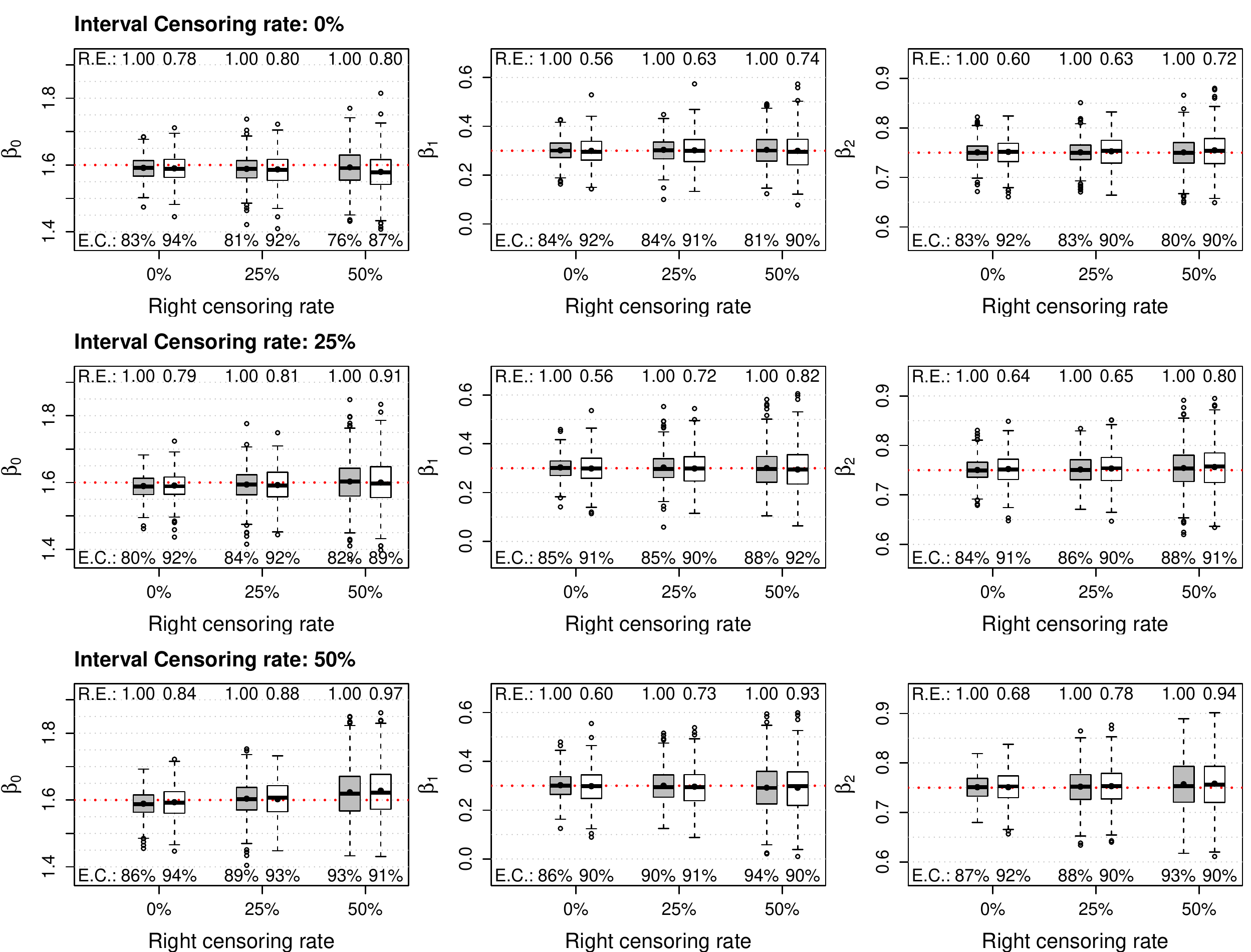}\\
  {\small Dispersion parameters ($n=250$)}\\
  \includegraphics[width=12.5cm]{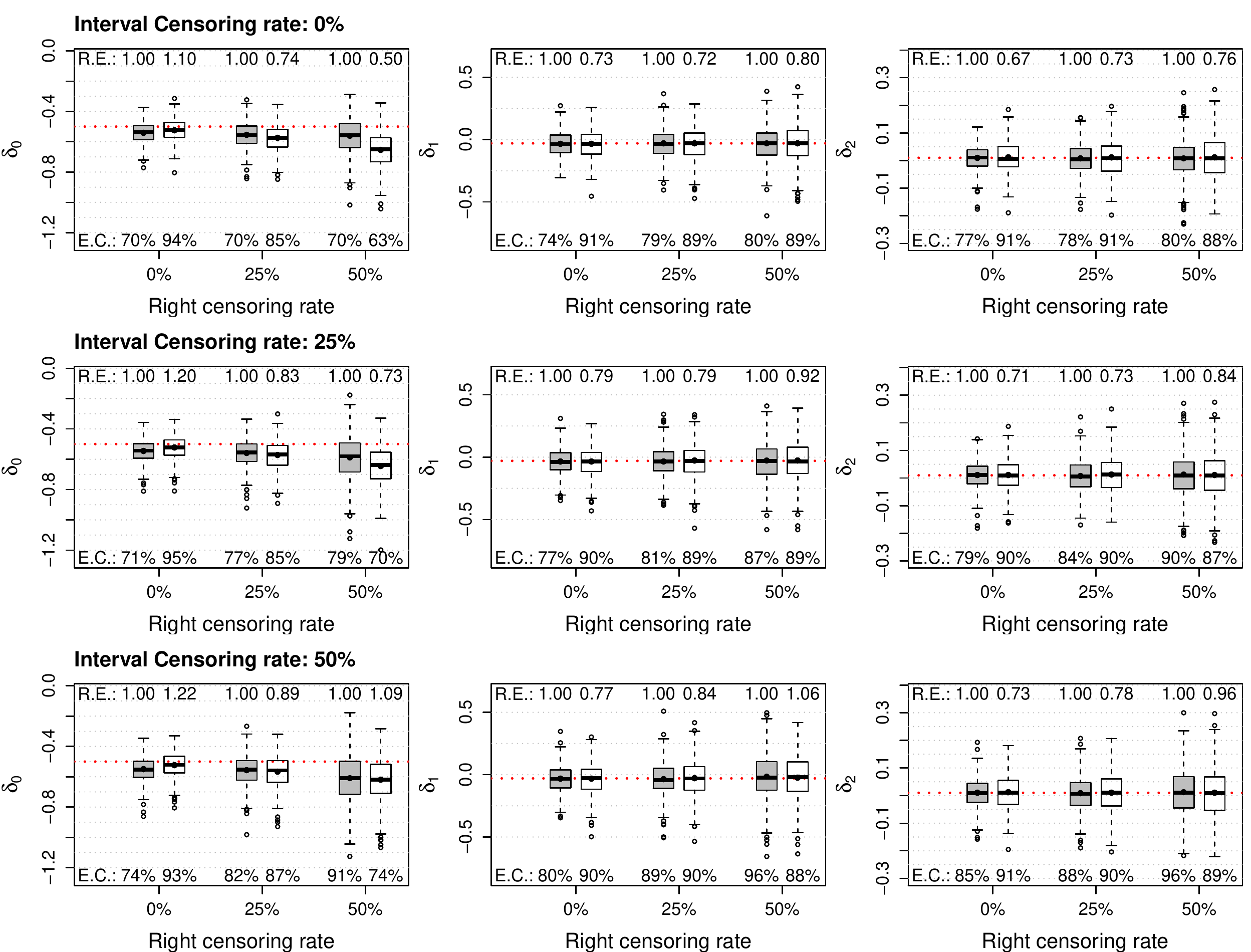}
\caption{Simulation study $(n=250)$: estimation of the regression parameters in
  the double additive location-scale model over $S=500$ replicates:
  boxplot of the point estimates under a nonparametric (grey) or Normal (white)
  error term, Relative Efficiency (R.E.) under the working Normality
  hypothesis, Effective Coverage (E.C.) of 95\% credible intervals.
} \label{SimulationRegrParameters-n250:Fig}
\end{figure}

The double additive location-scale model (DALSM) was fitted by
assuming a nonparametric (NP) or a Normal ($\cal N$) density for the
error term with 10 (=$L$) B-splines (associated to equidistant knots on
$(0,1)$) to reconstruct each of the additive terms and 20 (=$K$)
B-splines (associated to equidistant knots on $(-6,6)$) to estimate
the (log of the hazard function underlying the) nonnormal error
density. Figures \ref{SimulationRegrParameters-n1500:Fig},
\ref{SimulationRegrParameters-n500:Fig} and
\ref{SimulationRegrParameters-n250:Fig} report on the estimation of
the regression parameters $\pmb{\beta}$ and $\pmb{\delta}$ for each of
the three sample sizes for the nine possible combinations of right
and interval censoring rates. The boxplots inform us on the (sampling)
distribution of the parameter estimates (in grey for NP and white for
$\cal N$) over the $S=500$ replicates, R.E.~indicates the Relative
Efficiency (defined as the ratio of the mean squared errors) under a
working normality hypothesis (a value smaller than 1.0 suggesting than
the NP assumption is preferable), while E.C.~reports the Effective
Coverage of 95\% credible intervals (computed as
$\hat\theta\pm 1.96~\mathrm{s.e.}(\hat\theta)$).  Whatever the
considered sample size, the bias in the estimation of the regression
parameters is practically zero under the proposed NP approach, except
for the intercept $\beta_0$ in the location part when all data are
censored (with IC=RC=50\%) and for the intercept $\delta_0$ in the
dispersion part where negative biases increasing with the RC rate tend
to appear.  Larger biases appear for the intercepts under the same
circumstances when assuming Normality for the error term. In addition,
mean squared errors are always (resp.\,nearly always) markedly larger
under the Normality hypothesis when $n=1500$ (resp.~$n=500$ or $250$)
(as revealed by the reported R.E.~values below 1.00 under $\cal N$).
For settings with negligible biases and when $n=1500$, the effective
coverages of credible intervals are close to their nominal value 95\%
whatever the considered assumption on the error distribution,
suggesting that the standard errors were properly quantified and the
posterior distribution of the parameters close to normality.  When
$n=500$ and biases are negligible, the coverages of credible intervals
are satisfactory for the location parameters, but tend to be slightly
smaller than the nominal value for the dispersion parameters under the
NP hypothesis. When the sample size is small ($n=250$, as compared to
the model complexity and the amount of censoring), while efficiency
gains are still observable for the NP approach, the effective
coverages of credible intervals are nearly always below the results
achieved under the normality working hypothesis. Our results (not
shown here) indicates an under-estimation under NP of the posterior
standard deviation of the regression parameters when information is
sparse.

\begin{figure}\centering
\begin{tabular}{c}
  \includegraphics[width=12.cm]{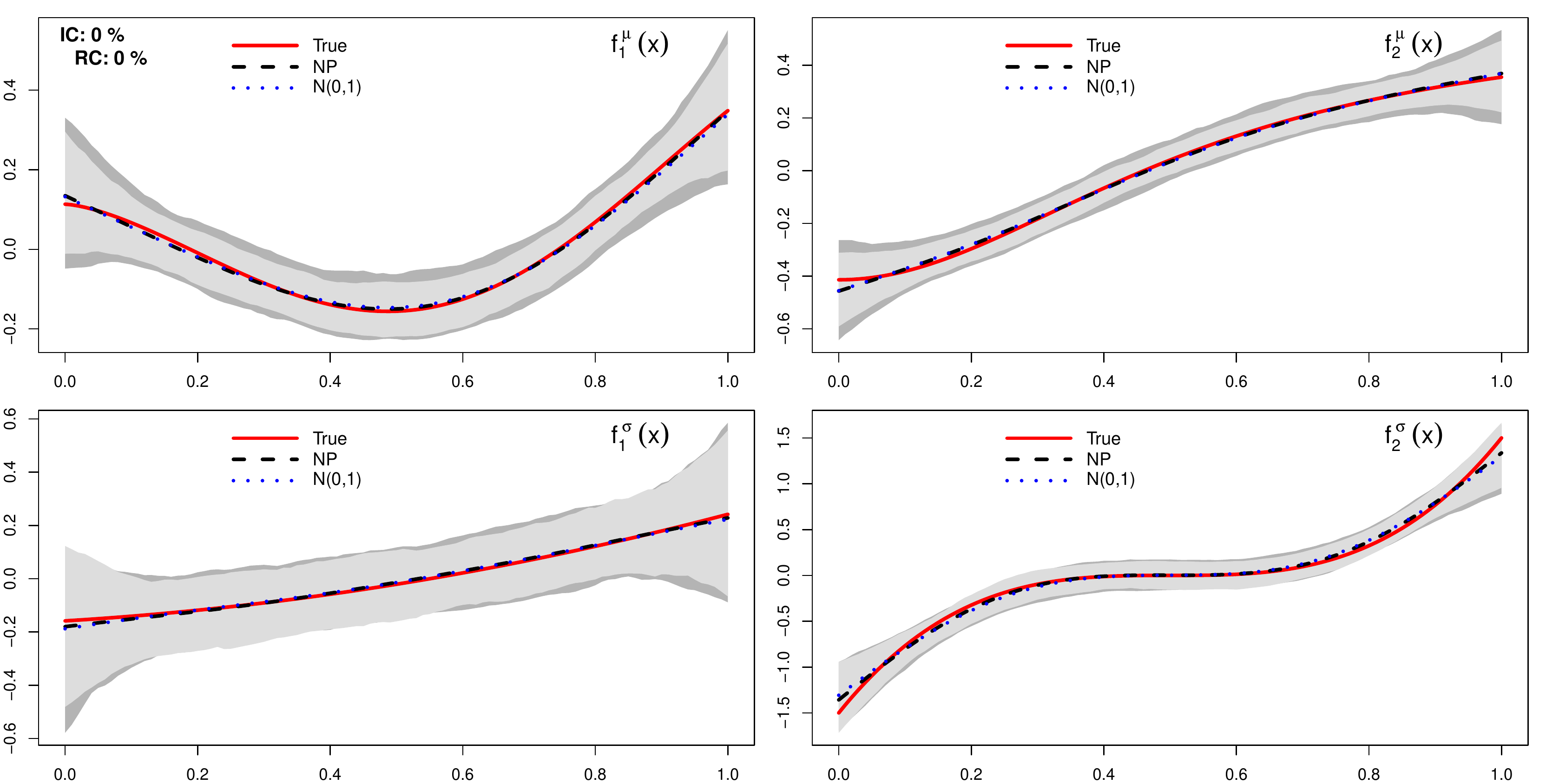}
  \\\\
  \includegraphics[width=12.cm]{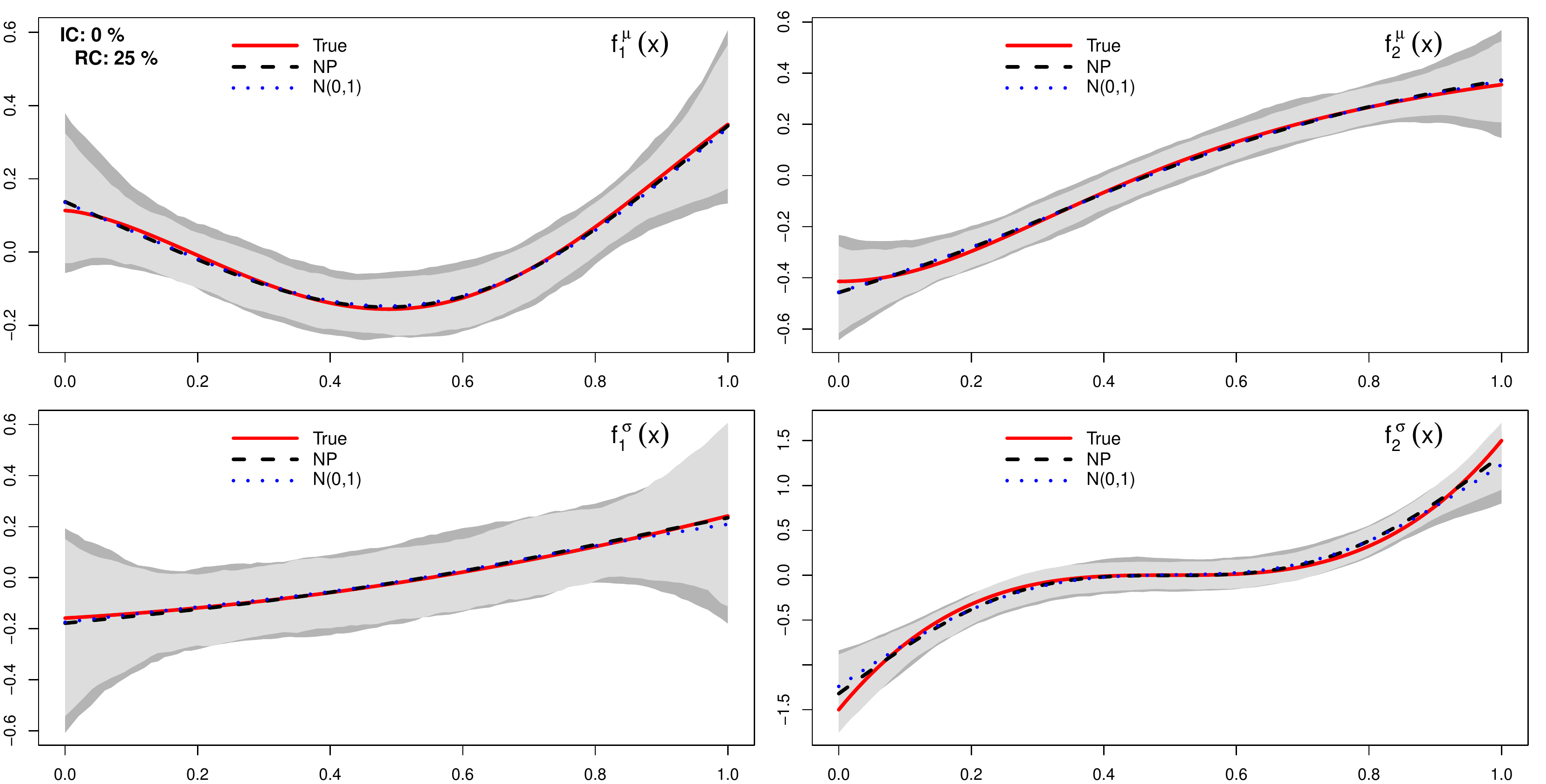}
  \\\\
  \includegraphics[width=12.cm]{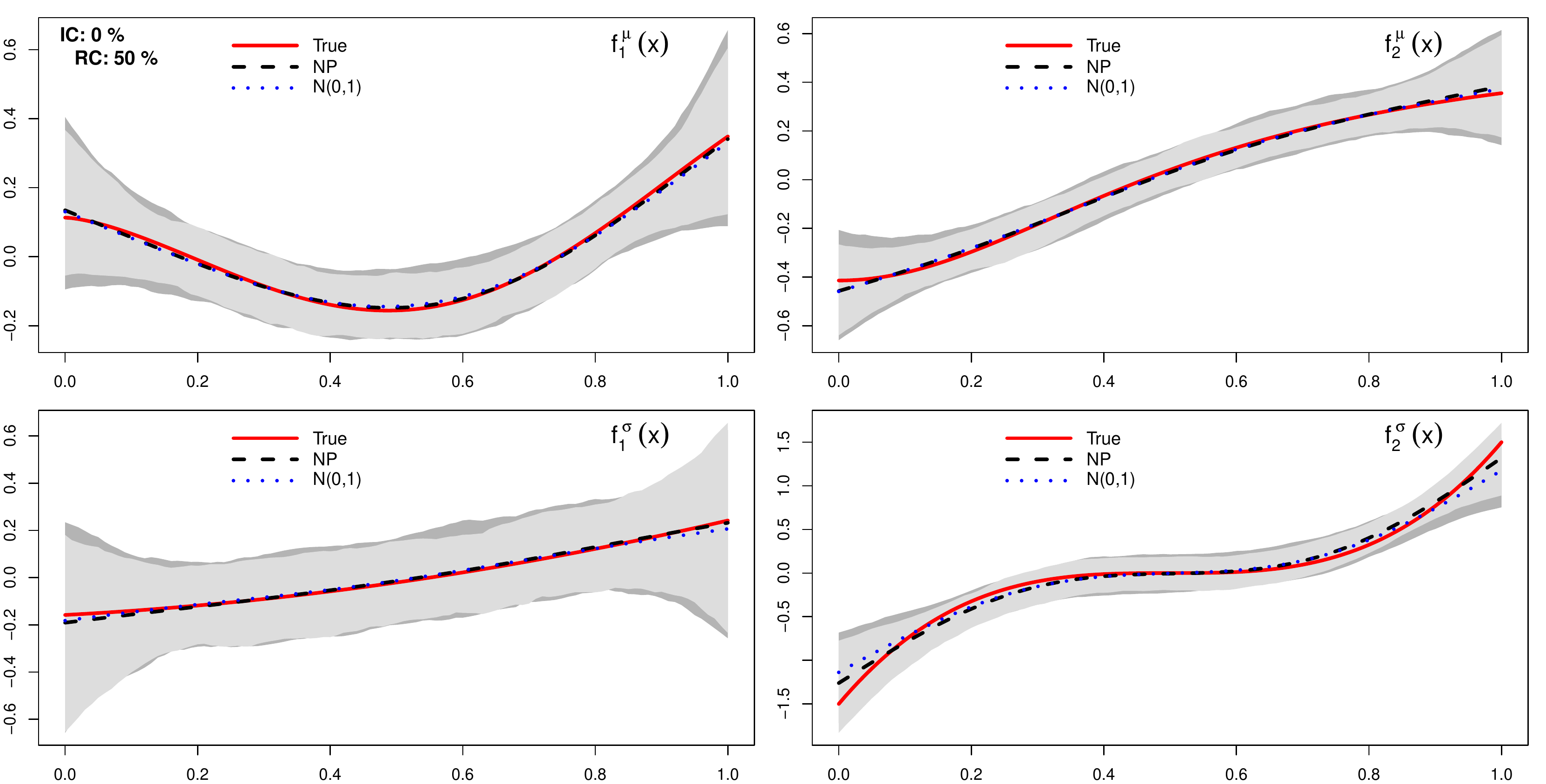}
\end{tabular}
\caption{Simulation study $(n=250)$: averaged estimated additive terms
  (over $S=500$ replicates) in the absence of interval censoring, but
  for increasing right censoring rates and by assuming a NP (dashed
  line) or a Normal (dotted line) error term. Envelopes (light grey:
  NP ; dark grey: Normal) result from consecutive intervals containing
  95\% of the $S$ estimates for $f^\mu_j(x)$ or $f^\sigma_j(x)$ with
  $x$ in $(0,1)$.  } \label{FittedAdditiveTermsN250-IC0:Fig}
\end{figure}

\begin{figure}\centering
\begin{tabular}{c}
  \includegraphics[width=12cm]{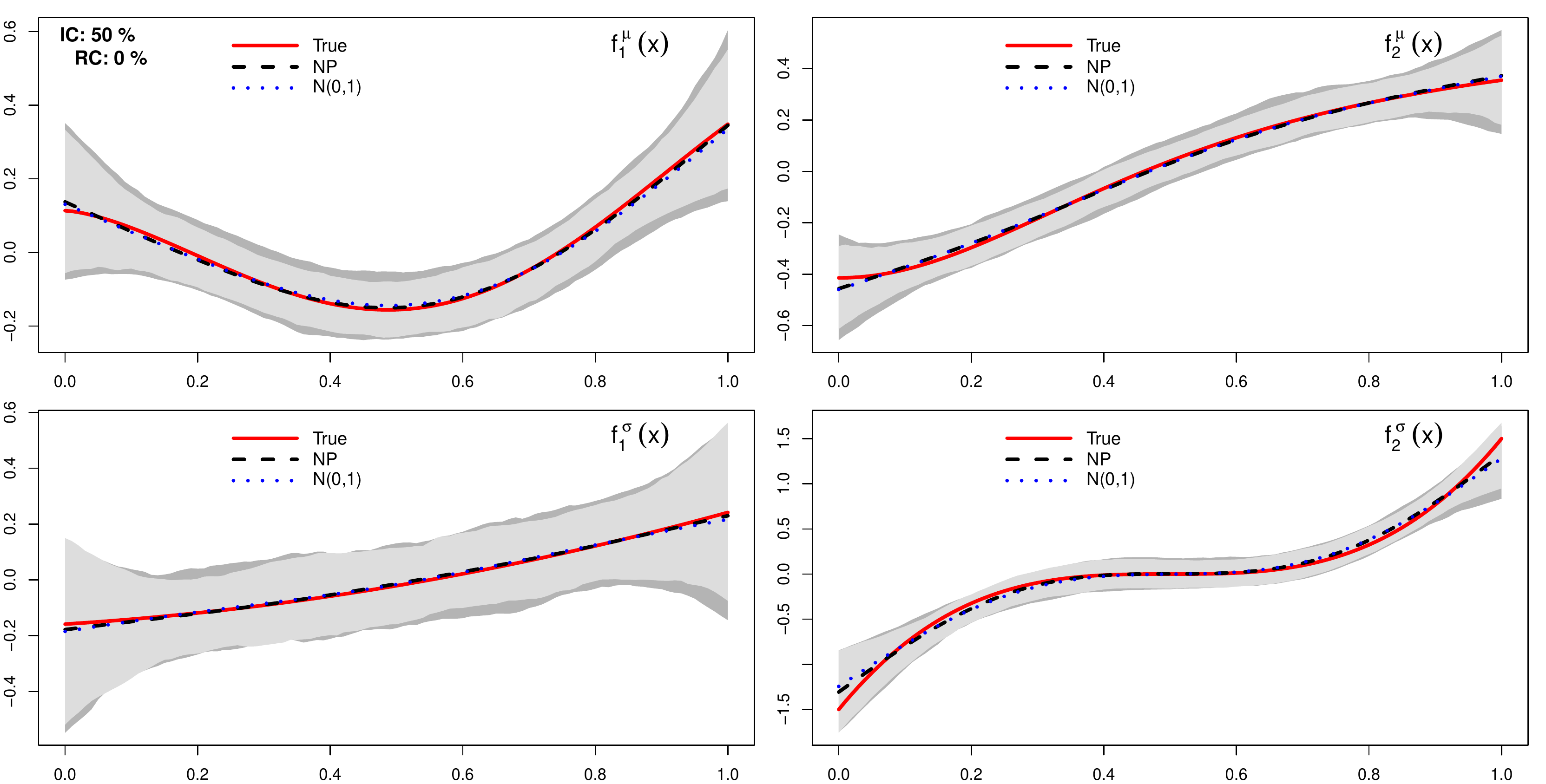}
  \\\\
  \includegraphics[width=12cm]{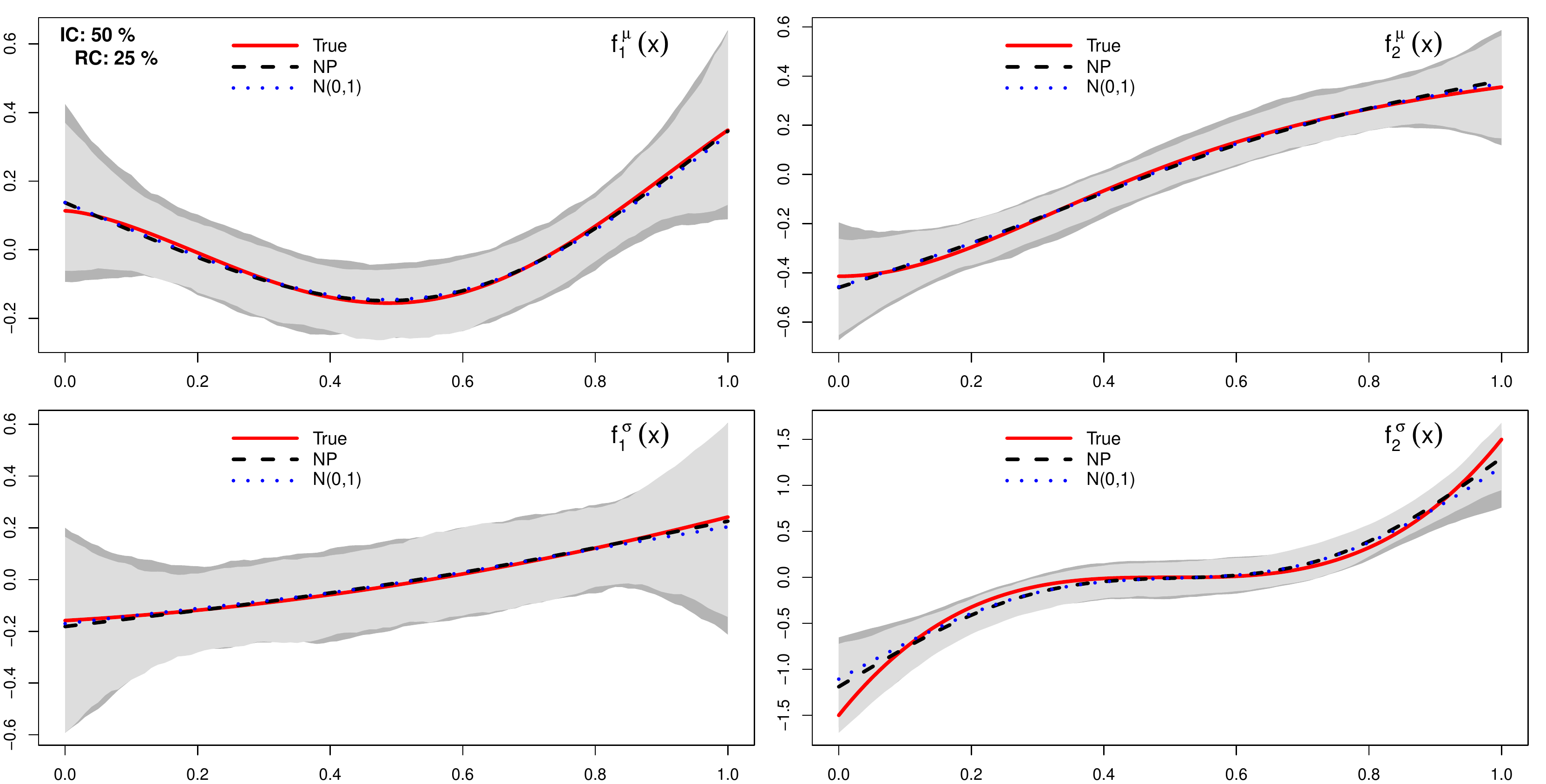}
  \\\\
  \includegraphics[width=12cm]{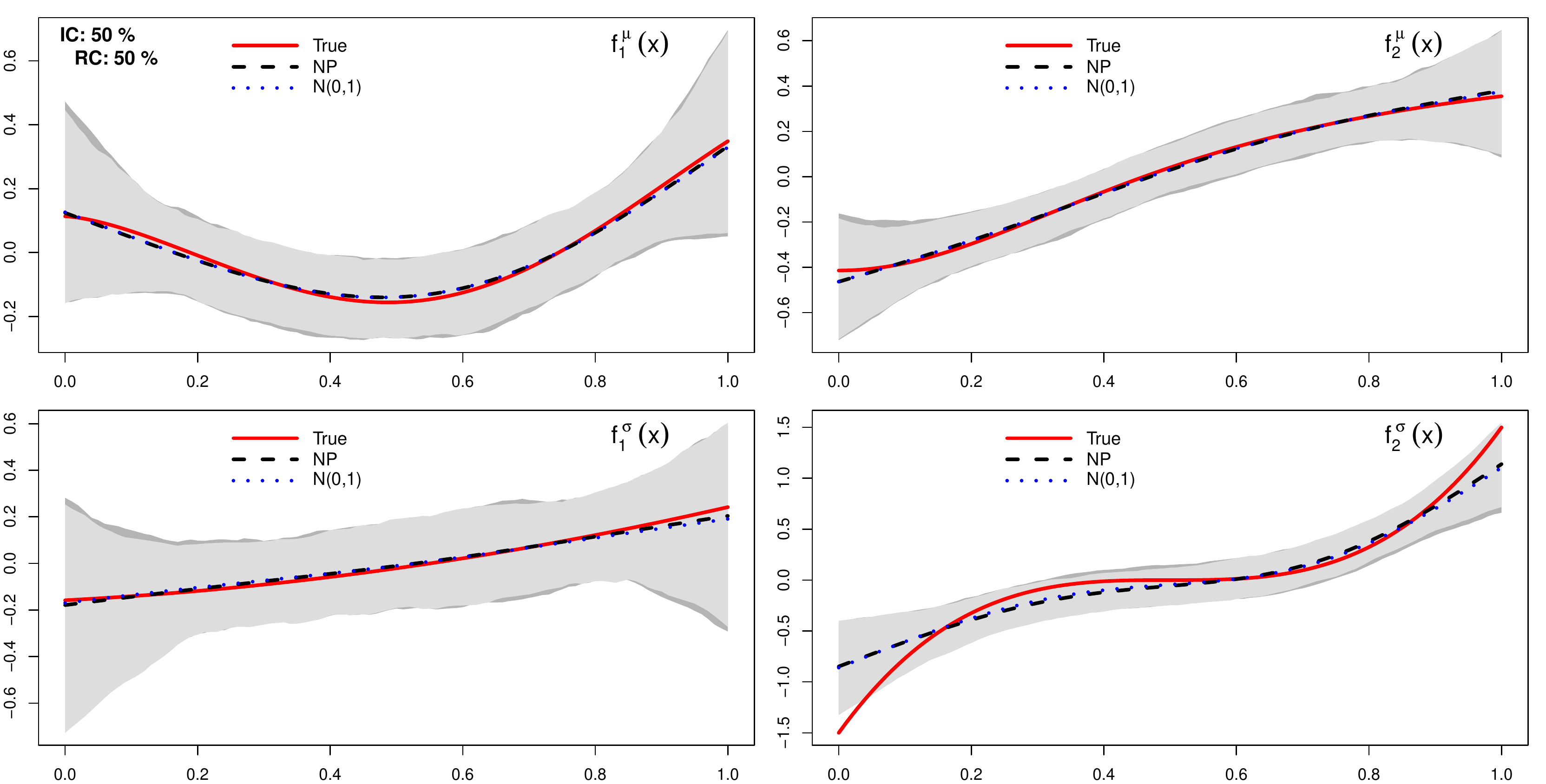}
\end{tabular}
\caption{Simulation study $(n=250)$: averaged estimated additive terms
  (over $S=500$ replicates) under a 50\% interval censoring rate
  combined with increasing right censoring rates and by
  assuming a NP (dashed line) or a Normal (dotted line) error
  term. Envelopes (light grey: NP ; dark grey: Normal) result from
  consecutive intervals containing 95\% of the $S$ estimates for
  $f^\mu_j(x)$ or $f^\sigma_j(x)$ with $x$ in $(0,1)$.
} \label{FittedAdditiveTermsN250-IC50:Fig}
\end{figure}

Report on the estimation of the additive terms can be found in Tables
\ref{Simulation-AdditiveTermsEstimation-n1500:Tab},
\ref{Simulation-AdditiveTermsEstimation-n500:Tab} and
\ref{Simulation-AdditiveTermsEstimation-n250:Tab}.  Whatever the
sample size and censoring rates, the absolute biases averaged over the
covariate support $(0,1)$ are very small, at the exception of
$f_2^\sigma(x)$ for values of $x$ close to zero when the sample size
is small ($n=250$) and the right and interval censoring rates are
large. Then, given the sparse information available, additive term
estimates naturally tend to be oversmoothed. It probably explains part
of the bias reported during the estimation of the intercept $\beta_0$
or $\delta_0$. This is illustrated in
Fig.~\ref{FittedAdditiveTermsN250-IC0:Fig} and
\ref{FittedAdditiveTermsN250-IC50:Fig} when the interval censoring
rate is 0\% or 50\%, respectively, for increasing right censoring
rates.  The wider dark grey envelope (connecting successive intervals
containing 95\% of the additive term estimates $f_j^\mu(x)$ or
$f_j^\sigma(x)$ over the $S$ replicates) also indicate that the
working Normality hypothesis for the error term yields less efficient
estimates than under the NP assumption (with light-grey
envelopes). This is confirmed numerically by the relative efficiency
values reported in the preceding tables.
The effective coverages of 95\% credible intervals for $f_j^\mu(x)$ or
$f_j^\sigma(x)$ averaged over the support $(0,1)$
of the covariate and the $S$ replicates are close to their nominal
values, except when information is sparse as it
naturally results in over-smoothing.

\begin{figure}\centering
  \includegraphics[width=13.5cm]{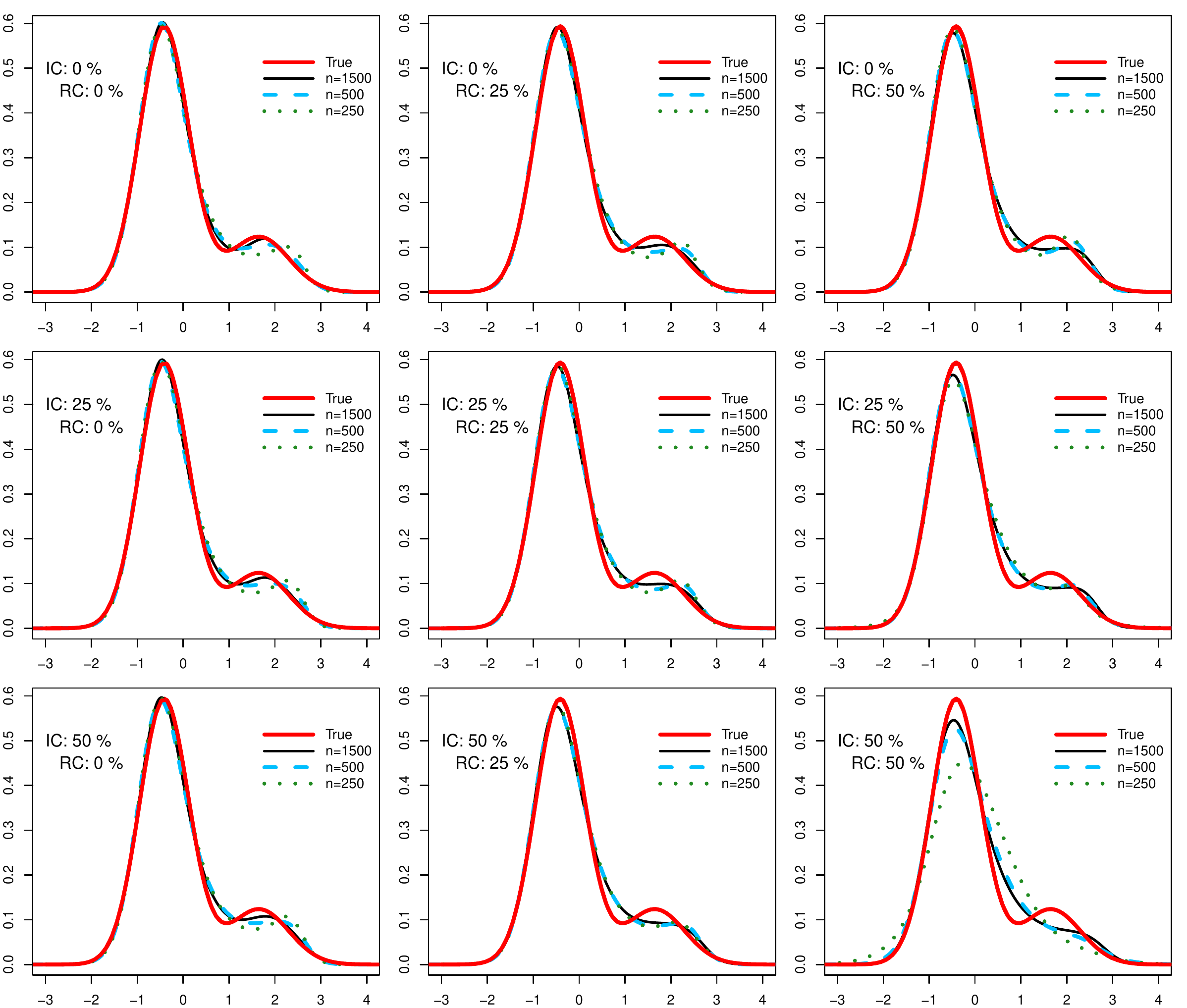}\\
  \caption{Simulation study: estimated error densities in the
    double additive location-scale model (averaged over the $S=500$
    replicates) using a NP error term for different combinations of sample sizes,
    right- (RC) and interval censoring (IC) rates.} \label{FittedErrorDensities:Fig}
\end{figure}

The estimates of the NP error density (averaged over the $S$
replicates) are given in Fig.~\ref{FittedErrorDensities:Fig} for
different combinations of right- and interval censoring rates.  When
the sample is large and in the absence of right censoring, the density
is very well estimated with an excellent performance of the selection
procedure for the underlying smoothness parameter
(cf.~Section
2.5.5). %\ref{SelectionPenaltyParmTau:Sec}).
Large right censoring rates have
an important negative effect on the quality of the reconstruction as
it reduces the ability to detect or position the second mode of the
target density. Combined with a large interval censoring rate and a
small sample size, it can even result in a right-skewed unimodal
average density estimate (see the dotted curve at the bottom right of
the figure) with the smallest component in the Normal mixture tending
to be flattened around its mode.

% %% Bibliography
% \bibliographystyle{Chicago}

% \bibliography{LocationScale}

\newpage
\pagestyle{empty} % To remove page numbers from here on
%%% ----------------------
%%% Simulation study: Figs
%%% ----------------------
%%% n = 1500

%%% n = 500

%%% n = 250

%% Additive terms
%%%%%%%%%%%%%%%%%

\spacingset{1} % DON'T change the spacing!
%%%%%%%%%%%%%%%%%%%%%%%%%%%%%%%%%%%%%
%%% n = 1500 - Location & Dispersion
%%%%%%%%%%%%%%%%%%%%%%%%%%%%%%%%%%%%%
\begin{table}%[ht]
\begin{center}
\rotatebox{90}{
\begin{minipage}{\textheight}
\begin{center}
\caption{\small Simulation study $(n=1500)$: estimation of the additive terms
  %in the \underline{conditional mean}
  in the double additive
  location-scale model over $S=500$ replicates
  for varying right censoring (RC) and interval censoring (IC) rates:
  Mean absolute bias, Root mean integrated squared error (RMISE),
  Relative Efficiency with an assumed Normal ($\Nc$) or nonparametric
  (NP) error term, Mean effective coverage of 95\% credible
  intervals.}
{%\small
{\small
\begin{tabular}{llcrrr@{\hspace{.6cm}}rrr@{\hspace{.6cm}}rrr@{\hspace{.7cm}}rrr}
  \hline
\multicolumn{2}{c}{$\bf n=1500$}           &       & \multicolumn{3}{c}{$f_1^\mu(x)$} & \multicolumn{3}{c}{$f_2^\mu(x)$}
                                  & \multicolumn{3}{c}{$f_1^\sigma(x)$} & \multicolumn{3}{c}{$f_2^\sigma(x)$}\\
 \cline{4-15}
IC &  \multicolumn{2}{r}{RC:}
                   & 0\%   & 25\%  & 50\%  & 0\%   & 25\%  & 50\%  &    0\%   & 25\%  & 50\%  & 0\%   & 25\%  & 50\% \\
  \hline
&  MA-Bias & NP    & 0.003 & 0.003 & 0.003 & 0.004 & 0.004 & 0.005 &   0.002 & 0.002 & 0.003 & 0.007 & 0.010 & 0.012\\
&          & $\Nc$ & 0.003 & 0.003 & 0.004 & 0.006 & 0.006 & 0.006 &   0.002 & 0.004 & 0.008 & 0.011 & 0.015 & 0.027\\
&  RMISE   & NP    & 0.015 & 0.017 & 0.021 & 0.016 & 0.018 & 0.021 &   0.025 & 0.029 & 0.036 & 0.036 & 0.041 & 0.050\\
0\%&       & $\Nc$ & 0.022 & 0.024 & 0.028 & 0.022 & 0.024 & 0.028 &   0.035 & 0.040 & 0.045 & 0.048 & 0.058 & 0.076\\
&Rel.Eff.  & NP    & 1.000 & 1.000 & 1.000 & 1.000 & 1.000 & 1.000 &   1.000 & 1.000 & 1.000 & 1.000 & 1.000 & 1.000\\
&          & $\Nc$ & 0.471 & 0.507 & 0.603 & 0.521 & 0.540 & 0.588 &   0.493 & 0.540 & 0.663 & 0.566 & 0.608 & 0.676\\
&Coverage  & NP    & 0.959 & 0.958 & 0.955 & 0.930 & 0.929 & 0.928 &   0.952 & 0.953 & 0.940 & 0.946 & 0.940 & 0.926\\
& 95\% CI  & $\Nc$ & 0.959 & 0.957 & 0.956 & 0.929 & 0.930 & 0.934 &   0.953 & 0.954 & 0.951 & 0.952 & 0.944 & 0.907\\
\hline
&  MA-Bias & NP    & 0.003 & 0.003 & 0.003 & 0.004 & 0.005 & 0.006 &   0.002 & 0.002 & 0.003 & 0.009 & 0.011 & 0.015\\
&          & $\Nc$ & 0.003 & 0.003 & 0.004 & 0.006 & 0.006 & 0.007 &   0.003 & 0.004 & 0.008 & 0.012 & 0.016 & 0.029\\
&  RMISE   & NP    & 0.017 & 0.020 & 0.026 & 0.017 & 0.020 & 0.025 &   0.026 & 0.032 & 0.040 & 0.038 & 0.046 & 0.059\\
25\%&      & $\Nc$ & 0.023 & 0.026 & 0.031 & 0.023 & 0.026 & 0.030 &   0.037 & 0.041 & 0.047 & 0.050 & 0.062 & 0.083\\
&Rel.Eff.  & NP    & 1.000 & 1.000 & 1.000 & 1.000 & 1.000 & 1.000 &   1.000 & 1.000 & 1.000 & 1.000 & 1.000 & 1.000\\
&          & $\Nc$ & 0.517 & 0.572 & 0.691 & 0.552 & 0.585 & 0.673 &   0.508 & 0.614 & 0.752 & 0.587 & 0.671 & 0.762\\
&Coverage  & NP    & 0.958 & 0.954 & 0.956 & 0.931 & 0.935 & 0.933 &   0.956 & 0.943 & 0.948 & 0.947 & 0.936 & 0.932\\
& 95\% CI  & $\Nc$ & 0.959 & 0.955 & 0.959 & 0.929 & 0.939 & 0.941 &   0.951 & 0.953 & 0.953 & 0.951 & 0.941 & 0.907\\
\hline
&  MA-Bias & NP    & 0.003 & 0.003 & 0.004 & 0.005 & 0.006 & 0.007 &   0.002 & 0.003 & 0.004 & 0.010 & 0.013 & 0.028\\
&          & $\Nc$ & 0.003 & 0.004 & 0.004 & 0.006 & 0.007 & 0.007 &   0.003 & 0.004 & 0.009 & 0.013 & 0.019 & 0.039\\
&  RMISE   & NP    & 0.019 & 0.023 & 0.032 & 0.019 & 0.023 & 0.031 &   0.029 & 0.036 & 0.045 & 0.042 & 0.052 & 0.080\\
50\%&      & $\Nc$ & 0.025 & 0.029 & 0.036 & 0.025 & 0.029 & 0.034 &   0.038 & 0.044 & 0.051 & 0.053 & 0.068 & 0.102\\
&Rel.Eff.  & NP    & 1.000 & 1.000 & 1.000 & 1.000 & 1.000 & 1.000 &   1.000 & 1.000 & 1.000 & 1.000 & 1.000 & 1.000\\
&          & $\Nc$ & 0.587 & 0.649 & 0.809 & 0.603 & 0.641 & 0.814 &   0.551 & 0.676 & 0.829 & 0.632 & 0.701 & 0.909\\
&Coverage  & NP    & 0.956 & 0.960 & 0.966 & 0.929 & 0.926 & 0.948 &   0.955 & 0.945 & 0.971 & 0.945 & 0.945 & 0.954\\
& 95\% CI  & $\Nc$ & 0.956 & 0.955 & 0.961 & 0.932 & 0.936 & 0.949 &   0.952 & 0.955 & 0.959 & 0.950 & 0.934 & 0.888\\
\hline
\end{tabular}
}
} \label{Simulation-AdditiveTermsEstimation-n1500:Tab}
\end{center}
\end{minipage}
}
\end{center}
\end{table}

%%%%%%%%%%%%%%%%%%%%%%%%%%%%%%%%%%%%%
%%% n = 500 - Location & Dispersion
%%%%%%%%%%%%%%%%%%%%%%%%%%%%%%%%%%%%%
\begin{table}%[ht]
\begin{center}
\rotatebox{90}{
\begin{minipage}{\textheight}
\begin{center}
\caption{\small Simulation study $(n=500)$: estimation of the additive terms
  %in the \underline{conditional mean}
  in the double additive location-scale model over $S=500$ replicates
  for varying right censoring (RC) and interval censoring (IC) rates:
  Mean absolute bias, Root mean integrated squared error (RMISE),
  Relative Efficiency with an assumed Normal ($\Nc$) or nonparametric
  (NP) error term, Mean effective coverage of 95\% credible
  intervals.}
{\small
\begin{tabular}{llcrrr@{\hspace{.6cm}}rrr@{\hspace{.6cm}}rrr@{\hspace{.7cm}}rrr}
  \hline
\multicolumn{2}{c}{$\bf n=500$}           &       & \multicolumn{3}{c}{$f_1^\mu(x)$} & \multicolumn{3}{c}{$f_2^\mu(x)$}
                                 & \multicolumn{3}{c}{$f_1^\sigma(x)$} & \multicolumn{3}{c}{$f_2^\sigma(x)$}\\
 \cline{4-15}
IC &  \multicolumn{2}{r}{RC:}
                   & 0\%   & 25\%  & 50\%  & 0\%   & 25\%  & 50\%  &    0\%   & 25\%  & 50\%  & 0\%   & 25\%  & 50\% \\
  \hline
&  MA-Bias & NP    & 0.004 & 0.004 & 0.004 & 0.005 & 0.005 & 0.007 &    0.002 & 0.003 & 0.004 & 0.018 & 0.021 & 0.025 \\
&          & $\Nc$ & 0.006 & 0.006 & 0.006 & 0.007 & 0.008 & 0.008 &    0.005 & 0.005 & 0.008 & 0.022 & 0.028 & 0.038 \\
&  RMISE   & NP    & 0.027 & 0.032 & 0.039 & 0.026 & 0.030 & 0.036 &    0.046 & 0.055 & 0.069 & 0.064 & 0.075 & 0.089 \\
0\%&       & $\Nc$ & 0.037 & 0.042 & 0.048 & 0.034 & 0.038 & 0.045 &    0.062 & 0.070 & 0.080 & 0.080 & 0.095 & 0.117 \\
&Rel.Eff.  & NP    & 1.000 & 1.000 & 1.000 & 1.000 & 1.000 & 1.000 &    1.000 & 1.000 & 1.000 & 1.000 & 1.000 & 1.000 \\
&          & $\Nc$ & 0.515 & 0.577 & 0.658 & 0.568 & 0.623 & 0.633 &    0.554 & 0.626 & 0.738 & 0.662 & 0.746 & 0.785 \\
&Coverage  & NP    & 0.944 & 0.935 & 0.919 & 0.922 & 0.911 & 0.906 &    0.918 & 0.910 & 0.900 & 0.899 & 0.883 & 0.882 \\
& 95\% CI  & $\Nc$ & 0.945 & 0.943 & 0.942 & 0.946 & 0.942 & 0.929 &    0.942 & 0.942 & 0.934 & 0.934 & 0.920 & 0.901 \\
\hline
&  MA-Bias & NP    & 0.004 & 0.004 & 0.004 & 0.005 & 0.006 & 0.009 &    0.003 & 0.002 & 0.006 & 0.020 & 0.024 & 0.036 \\
&          & $\Nc$ & 0.006 & 0.006 & 0.006 & 0.008 & 0.008 & 0.010 &    0.005 & 0.005 & 0.008 & 0.024 & 0.032 & 0.048 \\
&  RMISE   & NP    & 0.030 & 0.036 & 0.046 & 0.029 & 0.034 & 0.042 &    0.051 & 0.060 & 0.076 & 0.069 & 0.082 & 0.105 \\
25\%&      & $\Nc$ & 0.040 & 0.045 & 0.053 & 0.037 & 0.042 & 0.051 &    0.064 & 0.073 & 0.086 & 0.083 & 0.100 & 0.131 \\
&Rel.Eff.  & NP    & 1.000 & 1.000 & 1.000 & 1.000 & 1.000 & 1.000 &    1.000 & 1.000 & 1.000 & 1.000 & 1.000 & 1.000 \\
&          & $\Nc$ & 0.569 & 0.614 & 0.750 & 0.612 & 0.651 & 0.663 &    0.626 & 0.673 & 0.791 & 0.735 & 0.812 & 0.878 \\
&Coverage  & NP    & 0.940 & 0.932 & 0.930 & 0.917 & 0.912 & 0.926 &    0.905 & 0.921 & 0.929 & 0.890 & 0.885 & 0.903 \\
& 95\% CI  & $\Nc$ & 0.949 & 0.946 & 0.947 & 0.942 & 0.940 & 0.934 &    0.939 & 0.942 & 0.927 & 0.937 & 0.918 & 0.889 \\
\hline
&  MA-Bias & NP    & 0.004 & 0.005 & 0.006 & 0.006 & 0.008 & 0.010 &    0.003 & 0.003 & 0.007 & 0.021 & 0.028 & 0.060 \\
&          & $\Nc$ & 0.007 & 0.006 & 0.007 & 0.008 & 0.009 & 0.010 &    0.005 & 0.006 & 0.010 & 0.026 & 0.036 & 0.067 \\
&  RMISE   & NP    & 0.033 & 0.041 & 0.056 & 0.032 & 0.038 & 0.052 &    0.055 & 0.063 & 0.083 & 0.073 & 0.093 & 0.138 \\
50\%&      & $\Nc$ & 0.042 & 0.050 & 0.061 & 0.039 & 0.045 & 0.058 &    0.067 & 0.076 & 0.090 & 0.088 & 0.110 & 0.156 \\
&Rel.Eff.  & NP    & 1.000 & 1.000 & 1.000 & 1.000 & 1.000 & 1.000 &    1.000 & 1.000 & 1.000 & 1.000 & 1.000 & 1.000 \\
&          & $\Nc$ & 0.609 & 0.693 & 0.837 & 0.654 & 0.731 & 0.806 &    0.674 & 0.705 & 0.854 & 0.739 & 0.857 & 0.932 \\
&Coverage  & NP    & 0.942 & 0.940 & 0.963 & 0.921 & 0.931 & 0.956 &    0.909 & 0.939 & 0.968 & 0.898 & 0.906 & 0.939 \\
& 95\% CI  & $\Nc$ & 0.946 & 0.941 & 0.950 & 0.945 & 0.945 & 0.942 &    0.944 & 0.944 & 0.928 & 0.933 & 0.911 & 0.851 \\
\hline
\end{tabular}
} \label{Simulation-AdditiveTermsEstimation-n500:Tab}
\end{center}
\end{minipage}
}
\end{center}
\end{table}

%%%%%%%%%%%%%%%%%%%%%%%%%%%%%%%%%%%%%
%%% n = 250 - Location & Dispersion
%%%%%%%%%%%%%%%%%%%%%%%%%%%%%%%%%%%%%
\begin{table}%[ht]
\begin{center}
\rotatebox{90}{
\begin{minipage}{\textheight}
\begin{center}
\caption{\small Simulation study $(n=250)$: estimation of the additive terms
  %in the \underline{conditional mean}
  in the double additive location-scale model over $S=500$ replicates
  for varying right censoring (RC) and interval censoring (IC) rates:
  Mean absolute bias, Root mean integrated squared error (RMISE),
  Relative Efficiency with an assumed Normal ($\Nc$) or nonparametric
  (NP) error term, Mean effective coverage of 95\% credible
  intervals.}
{\small
\begin{tabular}{llcrrr@{\hspace{.6cm}}rrr@{\hspace{.6cm}}rrr@{\hspace{.7cm}}rrr}
  \hline
\multicolumn{2}{c}{$\bf n=250$}          &       & \multicolumn{3}{c}{$f_1^\mu(x)$} & \multicolumn{3}{c}{$f_2^\mu(x)$}
                                 & \multicolumn{3}{c}{$f_1^\sigma(x)$} & \multicolumn{3}{c}{$f_2^\sigma(x)$}\\
 \cline{4-15}
IC &  \multicolumn{2}{r}{RC:}
                   & 0\%   & 25\%  & 50\%  & 0\%   & 25\%  & 50\%  &    0\%   & 25\%  & 50\%  & 0\%   & 25\%  & 50\% \\
  \hline

&  MA-Bias & NP    & 0.006 & 0.006 & 0.007 & 0.007 & 0.008 & 0.010 &    0.006 & 0.005 & 0.008 & 0.028 & 0.036 & 0.053 \\
&          & $\Nc$ & 0.008 & 0.008 & 0.009 & 0.008 & 0.008 & 0.008 &    0.007 & 0.007 & 0.009 & 0.039 & 0.049 & 0.064 \\
&  RMISE   & NP    & 0.041 & 0.048 & 0.059 & 0.038 & 0.045 & 0.055 &    0.081 & 0.091 & 0.112 & 0.103 & 0.114 & 0.139 \\
0\%&       & $\Nc$ & 0.053 & 0.059 & 0.069 & 0.050 & 0.054 & 0.062 &    0.089 & 0.103 & 0.121 & 0.116 & 0.135 & 0.162 \\
&Rel.Eff.  & NP    & 1.000 & 1.000 & 1.000 & 1.000 & 1.000 & 1.000 &    1.000 & 1.000 & 1.000 & 1.000 & 1.000 & 1.000 \\
&          & $\Nc$ & 0.602 & 0.647 & 0.740 & 0.555 & 0.660 & 0.762 &    0.840 & 0.771 & 0.869 & 0.851 & 0.808 & 0.889 \\
&Coverage  & NP    & 0.879 & 0.875 & 0.869 & 0.855 & 0.855 & 0.841 &    0.803 & 0.821 & 0.844 & 0.767 & 0.803 & 0.806 \\
& 95\% CI  & $\Nc$ & 0.924 & 0.925 & 0.914 & 0.926 & 0.925 & 0.926 &    0.934 & 0.921 & 0.910 & 0.901 & 0.883 & 0.862 \\
\hline
&  MA-Bias & NP    & 0.005 & 0.007 & 0.008 & 0.008 & 0.010 & 0.011 &    0.006 & 0.005 & 0.009 & 0.031 & 0.044 & 0.074 \\
&          & $\Nc$ & 0.008 & 0.008 & 0.010 & 0.008 & 0.008 & 0.010 &    0.007 & 0.007 & 0.011 & 0.042 & 0.057 & 0.080 \\
&  RMISE   & NP    & 0.045 & 0.053 & 0.069 & 0.042 & 0.049 & 0.066 &    0.086 & 0.094 & 0.122 & 0.108 & 0.123 & 0.163 \\
25\%&      & $\Nc$ & 0.057 & 0.065 & 0.077 & 0.052 & 0.058 & 0.069 &    0.092 & 0.108 & 0.124 & 0.121 & 0.145 & 0.181 \\
&Rel.Eff.  & NP    & 1.000 & 1.000 & 1.000 & 1.000 & 1.000 & 1.000 &    1.000 & 1.000 & 1.000 & 1.000 & 1.000 & 1.000 \\
&          & $\Nc$ & 0.645 & 0.675 & 0.811 & 0.616 & 0.684 & 0.885 &    0.866 & 0.761 & 0.961 & 0.872 & 0.832 & 0.922 \\
&Coverage  & NP    & 0.885 & 0.884 & 0.904 & 0.864 & 0.876 & 0.896 &    0.808 & 0.870 & 0.906 & 0.786 & 0.831 & 0.852 \\
& 95\% CI  & $\Nc$ & 0.922 & 0.919 & 0.920 & 0.933 & 0.931 & 0.934 &    0.928 & 0.919 & 0.918 & 0.900 & 0.874 & 0.845 \\
\hline
&  MA-Bias & NP    & 0.006 & 0.007 & 0.012 & 0.009 & 0.011 & 0.010 &    0.005 & 0.006 & 0.011 & 0.036 & 0.057 & 0.111 \\
&          & $\Nc$ & 0.009 & 0.009 & 0.012 & 0.008 & 0.008 & 0.009 &    0.007 & 0.008 & 0.014 & 0.046 & 0.067 & 0.105 \\
&  RMISE   & NP    & 0.051 & 0.062 & 0.084 & 0.046 & 0.057 & 0.076 &    0.087 & 0.101 & 0.126 & 0.114 & 0.140 & 0.208 \\
50\%&      & $\Nc$ & 0.060 & 0.071 & 0.087 & 0.056 & 0.065 & 0.079 &    0.095 & 0.110 & 0.128 & 0.128 & 0.160 & 0.208 \\
&Rel.Eff.  & NP    & 1.000 & 1.000 & 1.000 & 1.000 & 1.000 & 1.000 &    1.000 & 1.000 & 1.000 & 1.000 & 1.000 & 1.000 \\
&          & $\Nc$ & 0.716 & 0.753 & 0.939 & 0.662 & 0.750 & 0.923 &    0.843 & 0.849 & 0.977 & 0.854 & 0.879 & 0.971 \\
&Coverage  & NP    & 0.890 & 0.909 & 0.948 & 0.875 & 0.901 & 0.957 &    0.845 & 0.912 & 0.973 & 0.823 & 0.863 & 0.913 \\
& 95\% CI  & $\Nc$ & 0.924 & 0.922 & 0.926 & 0.935 & 0.931 & 0.939 &    0.930 & 0.922 & 0.915 & 0.901 & 0.866 & 0.811 \\
\hline
\end{tabular}
} \label{Simulation-AdditiveTermsEstimation-n250:Tab}
\end{center}
\end{minipage}
}
\end{center}
\end{table}

\end{document}